\documentclass[a4paper,fleqn]{cas-sc}

\usepackage[authoryear,longnamesfirst]{natbib}

\def\tsc#1{\csdef{#1}{\textsc{\lowercase{#1}}\xspace}}
\tsc{COR}
\tsc{LLM}
\tsc{MCL}
\tsc{OPW}
\tsc{CSO}
\tsc{SES}

\usepackage[dvipsnames]{xcolor}

\usepackage[most]{tcolorbox}
\usepackage{todonotes}

\usepackage{soul}

\usepackage{multirow}
\usepackage{makecell}
\usepackage{array}
\usepackage{longtable}
\usepackage{supertabular}
\usepackage{color, colortbl}
\usepackage{caption}
\usepackage{subfig}
\usepackage{xspace}
\usepackage{enumitem}
\usepackage{multicol}

\usepackage{tikz}
\usetikzlibrary{arrows.meta, calc, fit, backgrounds, positioning, decorations.pathmorphing}
\usetikzlibrary{arrows.meta, calc, fit, backgrounds, positioning, shadings, shadows.blur}

\newcommand{\eg}{\emph{e.g.}\xspace}
\newcommand{\ie}{\emph{i.e.}\xspace}
\newcommand{\etc}{\emph{etc.}\@\xspace}

\begin{document}
\let\WriteBookmarks\relax
\def\floatpagepagefraction{1}
\def\textpagefraction{.001}

\shorttitle{Assessing flood-related wellbeing from public discourse in Ireland using large language model as judge, 2012--2025}

\shortauthors{R. Luo et~al.}


\title[mode=title]{Assessing flood-related wellbeing from public discourse in Ireland using large language model as judge, 2012--2025}

\tnotemark[1]

\author[1]{R\'ois\'in Luo}[]
\cormark[1]
\ead{{jiaolin.luo,karyn.morrissey}@universityofgalway.ie}

\affiliation[1]{
        organization={J.E. Cairnes School of Business \& Economics, University of Galway},
                country={Ireland}}


\author[1,2]{Karyn Morrissey}[]

\affiliation[2]{
                organization={Ryan Institute, University of Galway},
                country={Ireland}}


\cortext[cor1]{Corresponding author}

\begin{abstract}
Flooding is an environmental hazard that produces social and psychological consequences for exposed individuals, shaping threat appraisal, perceived coping capacity, access to social support, and confidence in institutional response. This paper studies flood-related wellbeing in Ireland from 2012 to 2025 using large-scale, unobtrusive public discourse from Meta Content Library. We develop a bespoke flood-wellbeing instrument organised with three domains: \emph{affective--cognitive distress appraisal}, \emph{diminished perceived efficacy and agency}, and \emph{perceived relational--institutional connectedness}. We then design a flood-wellbeing reasoning pipeline with large language models to extract multi-indicator, multi-domain flood-wellbeing measures from public discourse at scale.

The analysis reveals both stressor activation and social--ecological resilience mediation pathways in Irish flood discourse. Flood stressors directly activate affective--cognitive distress appraisal, diminished perceived efficacy and agency, and perceived relational--institutional connectedness. The analysis further identifies two major mediation pathways: an appraisal--efficacy pathway linking flood exposure to threat appraisal, affective burden, uncertainty, and perceived coping capacity; and a social-buffering pathway in which social and institutional protective resources are associated with perceived efficacy and appraisal of flood-related adversity. The paper contributes an unobtrusive discourse-based approach to flood-wellbeing research by introducing a bespoke instrument, identifying activation and mediation pathways in flood discourse, and producing a temporal dataset for studying wellbeing, adaptation, and resilience under recurrent flood exposure.
\end{abstract}

\begin{graphicalabstract}
\begin{center}
\resizebox{1\linewidth}{!}{
\begingroup
\definecolor{titlenavy}{HTML}{17305E}
\definecolor{panelblue}{HTML}{2660A4}
\definecolor{panelgreen}{HTML}{3C7D3E}
\definecolor{panelteal}{HTML}{1F7C8C}
\definecolor{panelpurple}{HTML}{5B3E9B}
\definecolor{borderblue}{HTML}{C2D6EE}
\definecolor{bordergreen}{HTML}{CADEC4}
\definecolor{borderteal}{HTML}{BFDCE0}
\definecolor{borderpurple}{HTML}{D3C9EC}
\definecolor{beLight}{HTML}{7FB46B}\definecolor{beDark}{HTML}{2F6B34}
\definecolor{doLight}{HTML}{3FA0B5}\definecolor{doDark}{HTML}{14687A}
\definecolor{cnLight}{HTML}{7A66C2}\definecolor{cnDark}{HTML}{3F2C82}
\definecolor{bodytext}{HTML}{2B3A4A}
\definecolor{arrowblue}{HTML}{2056C8}\definecolor{arrowgreen}{HTML}{2E8B57}
\definecolor{dbBlight}{HTML}{A9C8F0}\definecolor{dbBdark}{HTML}{2C62B8}
\definecolor{dbPlight}{HTML}{B9ABE6}\definecolor{dbPdark}{HTML}{5A45A0}
\definecolor{chTeal}{HTML}{2FA8A0}\definecolor{chBlue}{HTML}{3E78C8}
\definecolor{chViolet}{HTML}{7D5BC6}\definecolor{chGreen}{HTML}{5DAE4A}
\definecolor{chAmber}{HTML}{E8A33D}\definecolor{chSky}{HTML}{6FC3D9}
\definecolor{pagebg}{HTML}{FBFCFE}\definecolor{pagebg2}{HTML}{EEF3FA}
\def\softshadow#1{\fill[black, opacity=0.06] (0.04,-0.05) circle (#1);\fill[black, opacity=0.05] (0.025,-0.03) circle (#1);}

\def\heartbadge#1#2#3{%
\begin{scope}[shift={#1}]
  \softshadow{0.64}
  \fill[white] (0,0) circle (0.7);                       
  \shade[inner color=#2, outer color=#3] (0,0) circle (0.62);
  \draw[white,opacity=0.5,line width=0.5pt] (0,0) circle (0.62);
  \begin{scope}\clip (0,0) circle (0.62);
    \fill[white,opacity=0.18] (-0.05,0.18) ellipse (0.52 and 0.34);
    \fill[white,opacity=0.22] (-0.16,0.34) ellipse (0.30 and 0.16);
  \end{scope}
  \draw[white, line width=1.8pt, line cap=round, line join=round]
    (0,-0.30) .. controls (-0.37,-0.02) and (-0.43,0.34) .. (-0.21,0.38)
    .. controls (-0.07,0.41) and (0,0.22) .. (0,0.13)
    .. controls (0,0.22) and (0.07,0.41) .. (0.21,0.38)
    .. controls (0.43,0.34) and (0.37,-0.02) .. (0,-0.30) -- cycle;
\end{scope}}
\def\personbadge#1#2#3{%
\begin{scope}[shift={#1}]
  \softshadow{0.64}
  \fill[white] (0,0) circle (0.7);
  \shade[inner color=#2, outer color=#3] (0,0) circle (0.62);
  \draw[white,opacity=0.5,line width=0.5pt] (0,0) circle (0.62);
  \begin{scope}\clip (0,0) circle (0.62);
    \fill[white,opacity=0.18] (-0.05,0.18) ellipse (0.52 and 0.34);
    \fill[white,opacity=0.22] (-0.16,0.34) ellipse (0.30 and 0.16);
  \end{scope}
  \draw[white, line width=1.8pt] (0,0.24) circle (0.17);
  \draw[white, line width=1.8pt, line cap=round] (-0.30,-0.34) .. controls (-0.30,-0.04) and (0.30,-0.04) .. (0.30,-0.34);
\end{scope}}
\def\groupbadge#1#2#3{%
\begin{scope}[shift={#1}]
  \softshadow{0.64}
  \fill[white] (0,0) circle (0.7);
  \shade[inner color=#2, outer color=#3] (0,0) circle (0.62);
  \draw[white,opacity=0.5,line width=0.5pt] (0,0) circle (0.62);
  \begin{scope}\clip (0,0) circle (0.62);
    \fill[white,opacity=0.16] (-0.05,0.20) ellipse (0.52 and 0.34);
    \fill[white,opacity=0.20] (-0.16,0.36) ellipse (0.30 and 0.16);
  \end{scope}
  \draw[white, line width=1.4pt] (-0.28,0.18) circle (0.11);
  \draw[white, line width=1.4pt] (0.28,0.18) circle (0.11);
  \draw[white, line width=1.4pt, line cap=round] (-0.45,-0.20) .. controls (-0.45,0.02) and (-0.11,0.02) .. (-0.11,-0.20);
  \draw[white, line width=1.4pt, line cap=round] (0.11,-0.20) .. controls (0.11,0.02) and (0.45,0.02) .. (0.45,-0.20);
  \draw[white, line width=1.6pt] (0,0.25) circle (0.14);
  \draw[white, line width=1.6pt, line cap=round] (-0.24,-0.32) .. controls (-0.24,-0.02) and (0.24,-0.02) .. (0.24,-0.32);
\end{scope}}
\def\minigloss{\begin{scope}\clip (0,0) circle (0.42);
  \fill[white,opacity=0.16] (-0.04,0.13) ellipse (0.36 and 0.24);
  \fill[white,opacity=0.20] (-0.11,0.24) ellipse (0.20 and 0.11);\end{scope}}
\def\heartmini#1#2#3{\begin{scope}[shift={#1}]
  \fill[black,opacity=0.05] (0.03,-0.04) circle (0.43);
  \shade[inner color=#2, outer color=#3] (0,0) circle (0.42);
  \minigloss
  \draw[white, line width=1.3pt, line cap=round, line join=round]
    (0,-0.20) .. controls (-0.25,-0.01) and (-0.29,0.23) .. (-0.14,0.26)
    .. controls (-0.05,0.28) and (0,0.15) .. (0,0.09)
    .. controls (0,0.15) and (0.05,0.28) .. (0.14,0.26)
    .. controls (0.29,0.23) and (0.25,-0.01) .. (0,-0.20) -- cycle;\end{scope}}
\def\personmini#1#2#3{\begin{scope}[shift={#1}]
  \fill[black,opacity=0.05] (0.03,-0.04) circle (0.43);
  \shade[inner color=#2, outer color=#3] (0,0) circle (0.42);
  \minigloss
  \draw[white, line width=1.3pt] (0,0.16) circle (0.115);
  \draw[white, line width=1.3pt, line cap=round] (-0.20,-0.23) .. controls (-0.20,-0.03) and (0.20,-0.03) .. (0.20,-0.23);\end{scope}}
\def\groupmini#1#2#3{\begin{scope}[shift={#1}]
  \fill[black,opacity=0.05] (0.03,-0.04) circle (0.43);
  \shade[inner color=#2, outer color=#3] (0,0) circle (0.42);
  \minigloss
  \draw[white, line width=1.0pt] (-0.19,0.12) circle (0.075);
  \draw[white, line width=1.0pt] (0.19,0.12) circle (0.075);
  \draw[white, line width=1.0pt, line cap=round] (-0.30,-0.14) .. controls (-0.30,0.01) and (-0.08,0.01) .. (-0.08,-0.14);
  \draw[white, line width=1.0pt, line cap=round] (0.08,-0.14) .. controls (0.08,0.01) and (0.30,0.01) .. (0.30,-0.14);
  \draw[white, line width=1.1pt] (0,0.17) circle (0.095);
  \draw[white, line width=1.1pt, line cap=round] (-0.16,-0.22) .. controls (-0.16,-0.01) and (0.16,-0.01) .. (0.16,-0.22);\end{scope}}

\def\stormcloud#1#2{%
\begin{scope}[shift={#1}, scale=#2]
  \fill[black,opacity=0.05] (0.05,0.25) ellipse (0.95 and 0.5);
  \shade[top color=gray!28, bottom color=gray!55]
    (-0.7,0.12) .. controls (-1.05,0.12) and (-1.02,0.62) .. (-0.58,0.62)
    .. controls (-0.5,0.95) and (0.16,0.99) .. (0.34,0.66)
    .. controls (0.7,0.8) and (0.98,0.4) .. (0.6,0.16) -- cycle;
  \shade[top color=gray!45, bottom color=gray!80]
    (-0.5,0.0) .. controls (-0.82,0.0) and (-0.8,0.44) .. (-0.42,0.44)
    .. controls (-0.36,0.72) and (0.2,0.75) .. (0.34,0.48)
    .. controls (0.64,0.6) and (0.86,0.26) .. (0.54,0.06) -- cycle;
  \foreach \x in {-0.5,-0.18,0.16,0.46}{\draw[NavyBlue!55, line width=1.2pt, line cap=round] (\x,-0.1)--(\x-0.14,-0.46);}
  \fill[orange!90, draw=orange!60!red, line width=0.4pt] (0.0,-0.06)--(-0.16,-0.44)--(-0.01,-0.44)--(-0.13,-0.8)--(0.18,-0.34)--(0.02,-0.34)--cycle;
\end{scope}}
\def\greycloud#1#2{%
\begin{scope}[shift={#1}, scale=#2]
  \fill[black,opacity=0.04] (0.05,0.2) ellipse (0.9 and 0.45);
  \shade[top color=gray!25, bottom color=gray!50]
    (-0.7,0.12) .. controls (-1.05,0.12) and (-1.02,0.62) .. (-0.58,0.62)
    .. controls (-0.5,0.95) and (0.16,0.99) .. (0.34,0.66)
    .. controls (0.7,0.8) and (0.98,0.4) .. (0.6,0.16) -- cycle;
  \shade[top color=gray!38, bottom color=gray!68]
    (-0.5,0.0) .. controls (-0.82,0.0) and (-0.8,0.44) .. (-0.42,0.44)
    .. controls (-0.36,0.72) and (0.2,0.75) .. (0.34,0.48)
    .. controls (0.64,0.6) and (0.86,0.26) .. (0.54,0.06) -- cycle;
  \foreach \x in {-0.45,-0.1,0.28,0.5}{\draw[NavyBlue!45, line width=1pt, line cap=round] (\x,-0.06)--(\x-0.12,-0.4);}
\end{scope}}

\def\database#1#2#3#4{%
\begin{scope}[shift={#1}, scale=#2]
  \fill[black,opacity=0.06] (0.05,-0.55) ellipse (0.62 and 0.16);
  \foreach \dy in {-0.6,-0.2,0.2}{
    \shade[left color=#4, right color=#3, draw=#4, line width=0.5pt]
      (-0.6,\dy) -- (-0.6,\dy+0.38)
      arc[start angle=180,end angle=360,x radius=0.6,y radius=0.18] -- (0.6,\dy)
      arc[start angle=0,end angle=-180,x radius=0.6,y radius=0.18] -- cycle;}
  \shade[left color=#4, right color=#3, draw=#4, line width=0.5pt]
    (-0.6,0.58) arc[start angle=180,end angle=360,x radius=0.6,y radius=0.18]
    arc[start angle=0,end angle=180,x radius=0.6,y radius=0.18];
  \fill[white, opacity=0.55] (-0.18,0.66) ellipse (0.30 and 0.10);
  \fill[white, opacity=0.30] (0.10,0.62) ellipse (0.16 and 0.05);
\end{scope}}

\def\brainnet#1#2{%
\begin{scope}[shift={#1}, scale=#2, line cap=round, line join=round]
  \draw[panelgreen, line width=1.4pt]
    (0,0.5) .. controls (-0.45,0.72) and (-0.85,0.42) .. (-0.78,0.05)
    .. controls (-1.0,-0.22) and (-0.62,-0.55) .. (-0.32,-0.45)
    .. controls (-0.18,-0.62) and (0.18,-0.62) .. (0.32,-0.45)
    .. controls (0.62,-0.55) and (1.0,-0.22) .. (0.78,0.05)
    .. controls (0.85,0.42) and (0.45,0.72) .. (0,0.5);
  \draw[panelgreen, line width=1.1pt] (0,0.5)--(0,-0.5);
  \draw[panelgreen!75, line width=0.7pt] (-0.5,0.22)--(-0.15,0.34) (0.2,0.26)--(0.5,0.06) (-0.28,-0.08)--(0.04,0.0) (0.04,0.0)--(0.32,-0.24);
  \foreach \p in {(-0.5,0.22),(0.2,0.26),(-0.28,-0.08),(0.32,-0.24),(-0.15,0.34),(0.5,0.06),(0.04,0.0)}{\fill[panelgreen]\p circle(0.045);}
\end{scope}}

\def\charticon#1#2{%
\begin{scope}[shift={#1}, scale=#2, line cap=round, line join=round]
  \shade[bottom color=chSky, top color=chSky!55] (-0.95,-0.44) rectangle ++(0.13,0.28);
  \shade[bottom color=chTeal, top color=chTeal!55] (-0.77,-0.44) rectangle ++(0.13,0.44);
  \shade[bottom color=chBlue, top color=chBlue!55] (-0.59,-0.44) rectangle ++(0.13,0.62);
  \shade[bottom color=chViolet, top color=chViolet!55] (-0.41,-0.44) rectangle ++(0.13,0.80);
  \draw[gray!40, line width=0.5pt] (-0.99,-0.44) -- (-0.22,-0.44);
  \fill[chBlue] (0.520,0.040) -- (0.520,0.460) arc[start angle=90.00, end angle=-10.80, radius=0.420] -- cycle;
  \fill[chViolet] (0.520,0.040) -- (0.933,-0.039) arc[start angle=-10.80, end angle=-90.00, radius=0.420] -- cycle;
  \fill[chTeal] (0.520,0.040) -- (0.520,-0.380) arc[start angle=-90.00, end angle=-162.00, radius=0.420] -- cycle;
  \fill[chGreen] (0.520,0.040) -- (0.121,-0.090) arc[start angle=-162.00, end angle=-219.60, radius=0.420] -- cycle;
  \fill[chAmber] (0.520,0.040) -- (0.196,0.308) arc[start angle=-219.60, end angle=-270.00, radius=0.420] -- cycle;
  \fill[white] (0.520,0.040) circle (0.15);
\end{scope}}

\def\peoplehill#1#2{%
\begin{scope}[shift={#1}, scale=#2]
  \shade[top color=panelgreen!45, bottom color=panelgreen!85] (-0.6,-0.05) .. controls (-0.2,0.16) and (0.2,0.16) .. (0.6,-0.05) -- (0.6,-0.32) -- (-0.6,-0.32) -- cycle;
  \foreach \x in {-0.3,0,0.3}{\fill[panelteal] (\x,0.2) circle (0.085); \fill[panelteal] (\x-0.1,-0.06) rectangle (\x+0.1,0.14);}
\end{scope}}

\def\chatbubble#1#2#3#4#5{
}
\def\speechbubbles#1#2{%
\begin{scope}[shift={#1}, scale=#2]
  \begin{scope}[shift={(0.34,0.42)}]
    \fill[black,opacity=0.07] (0.02,-0.05) ellipse (0.5 and 0.34);
    \shade[top color=panelblue!85, bottom color=panelblue]
      (-0.46,0.0) .. controls (-0.46,0.30) and (-0.30,0.40) .. (0.0,0.40)
      .. controls (0.30,0.40) and (0.46,0.30) .. (0.46,0.0)
      .. controls (0.46,-0.22) and (0.30,-0.32) .. (0.06,-0.34)
      .. controls (0.06,-0.34) and (0.0,-0.46) .. (-0.10,-0.52)  
      .. controls (-0.10,-0.40) and (-0.16,-0.34) .. (-0.24,-0.33)
      .. controls (-0.36,-0.30) and (-0.46,-0.20) .. (-0.46,0.0) -- cycle;
    \foreach \x in {-0.20,0.0,0.20}{\fill[white] (\x,0.04) circle (0.052);}
  \end{scope}
  \begin{scope}[shift={(-0.30,0.06)}]
    \fill[black,opacity=0.06] (0.02,-0.05) ellipse (0.46 and 0.32);
    \shade[top color=white, bottom color=blue!8]
      (-0.42,0.0) .. controls (-0.42,0.28) and (-0.27,0.37) .. (0.0,0.37)
      .. controls (0.27,0.37) and (0.42,0.28) .. (0.42,0.0)
      .. controls (0.42,-0.20) and (0.27,-0.30) .. (0.06,-0.31)
      .. controls (0.10,-0.40) and (0.06,-0.48) .. (-0.04,-0.52)  
      .. controls (-0.12,-0.42) and (-0.18,-0.32) .. (-0.26,-0.30)
      .. controls (-0.34,-0.28) and (-0.42,-0.18) .. (-0.42,0.0) -- cycle;
    \draw[panelblue!30, line width=0.6pt]
      (-0.42,0.0) .. controls (-0.42,0.28) and (-0.27,0.37) .. (0.0,0.37)
      .. controls (0.27,0.37) and (0.42,0.28) .. (0.42,0.0)
      .. controls (0.42,-0.20) and (0.27,-0.30) .. (0.06,-0.31)
      .. controls (0.10,-0.40) and (0.06,-0.48) .. (-0.04,-0.52)
      .. controls (-0.12,-0.42) and (-0.18,-0.32) .. (-0.26,-0.30)
      .. controls (-0.34,-0.28) and (-0.42,-0.18) .. (-0.42,0.0) -- cycle;
    \foreach \x in {-0.18,0.0,0.18}{\fill[panelblue!70] (\x,0.04) circle (0.048);}
  \end{scope}
\end{scope}}
\def\housei#1#2{%
\begin{scope}[shift={#1}, scale=#2, line join=round]
  \fill[panelblue!80] (-0.34,0.05) -- (0,0.38) -- (0.34,0.05) -- cycle;
  \fill[white, draw=panelblue!45, line width=0.5pt] (-0.25,-0.3) rectangle (0.25,0.08);
  \fill[panelblue!30] (-0.08,-0.3) rectangle (0.08,0.02);
\end{scope}}
\def\floodwater#1#2{%
\begin{scope}[shift={#1}, scale=#2]
  \foreach \y/\o in {0/0.95,0.16/0.78,0.32/0.6,0.48/0.42,0.62/0.28}{
    \shade[top color=panelblue!65, bottom color=panelblue, opacity=\o]
      (-1.5,\y-0.1) .. controls (-0.9,\y+0.08) and (-0.5,\y-0.1) .. (0,\y)
      .. controls (0.5,\y+0.1) and (0.9,\y-0.08) .. (1.5,\y-0.03)
      -- (1.5,\y-0.42) -- (-1.5,\y-0.42) -- cycle;}
\end{scope}}
\def\treei#1#2{%
\begin{scope}[shift={#1}, scale=#2]
  \draw[brown!55!black, line width=1.1pt] (0,-0.22)--(0,0.05);
  \shade[inner color=panelgreen!70, outer color=panelgreen] (0,0.12) ellipse (0.17 and 0.28);
  \fill[panelgreen!50,opacity=0.6] (-0.06,0.2) ellipse (0.08 and 0.13);
\end{scope}}

\def\rainfield#1#2{%
\begin{scope}[shift={#1}, scale=#2]
  \foreach \x in {-2.0,-1.7,-1.4,-1.1,-0.8,-0.5,-0.2,0.1,0.4,0.7,1.0,1.3,1.6,1.9}{
    \pgfmathsetmacro{\jit}{0.12*sin(\x*220)}
    \draw[panelblue!42, line width=0.7pt, line cap=round, opacity=0.55]
      (\x+\jit,0.0)--(\x+\jit-0.16,-0.5);
  }
  \foreach \x in {-1.85,-1.25,-0.65,-0.05,0.55,1.15,1.75}{
    \draw[panelblue!55, line width=0.9pt, line cap=round, opacity=0.7]
      (\x,0.28)--(\x-0.18,-0.28);
  }
\end{scope}}
\def\floodhouse#1#2#3{
\begin{scope}[shift={#1}, scale=#2, line join=round]
  \fill[panelblue!90, opacity=0.18] (-0.02,#3-0.02) ellipse (0.46 and 0.07);
  \shade[top color=panelblue, bottom color=panelblue!75]
    (-0.40,0.04) -- (0,0.44) -- (0.40,0.04) -- cycle;
  \draw[panelblue!55,line width=0.4pt] (0,0.44)--(0,0.04);
  \fill[white,opacity=0.26] (-0.18,0.18) -- (0,0.40) -- (-0.04,0.10) -- cycle;
  \shade[left color=white, right color=blue!12, draw=panelblue!50, line width=0.5pt]
    (-0.30,-0.34) rectangle (0.30,0.06);
  \fill[panelblue!28] (-0.10,-0.12) rectangle (0.10,0.0);
  \draw[panelblue!45,line width=0.4pt] (0,-0.12)--(0,0.0) (-0.10,-0.06)--(0.10,-0.06);
  \begin{scope}\clip (-0.5,#3) rectangle (0.5,#3-0.36);
    \shade[top color=panelblue!55, bottom color=panelblue!20, opacity=0.30]
      (-0.40,2*#3-0.04) -- (0,2*#3-0.44) -- (0.40,2*#3-0.04) -- cycle;
    \fill[white,opacity=0.10] (-0.30,2*#3-0.06) rectangle (0.30,2*#3+0.34);
  \end{scope}
\end{scope}}
\def\floodwaterB#1#2{%
\begin{scope}[shift={#1}, scale=#2]
  \foreach \y/\o/\fm in {0.00/0.95/1,0.20/0.80/1,0.40/0.62/0,0.58/0.45/0,0.74/0.30/0}{
    \shade[top color=panelblue!60, bottom color=panelblue, opacity=\o]
      (-1.62,\y-0.10) .. controls (-1.0,\y+0.10) and (-0.55,\y-0.12) .. (0,\y)
      .. controls (0.55,\y+0.12) and (1.0,\y-0.10) .. (1.62,\y-0.02)
      -- (1.62,\y-0.50) -- (-1.62,\y-0.50) -- cycle;
    \ifnum\fm=1
      \draw[white,opacity=0.30,line width=0.8pt,line cap=round]
        (-1.30,\y-0.05) .. controls (-0.8,\y+0.07) and (-0.4,\y-0.07) .. (0.05,\y+0.01)
        .. controls (0.5,\y+0.08) and (0.9,\y-0.05) .. (1.30,\y+0.01);
    \fi}
  \foreach \gx/\gy in {-1.0/0.10,-0.2/0.30,0.7/0.12,1.2/0.26,-0.6/0.42}{
    \fill[white,opacity=0.5] (\gx,\gy) circle (0.018);}
\end{scope}}

\noindent\hspace*{-4.4cm}\begin{tikzpicture}[font=\sffamily,
  panel/.style={rounded corners=12pt, line width=1.1pt},
  badge/.style={circle, text=white, font=\sffamily\bfseries, minimum size=0.66cm, inner sep=0pt},
  dompill/.style={rounded corners=10pt, draw=#1, fill=white, line width=1pt, minimum width=4.7cm, minimum height=1.7cm},
  pathbox/.style={rounded corners=9pt, draw=borderteal, fill=blue!2, line width=0.9pt}]
\shade[top color=pagebg, bottom color=pagebg2, rounded corners=16pt] (0.20,0.46) rectangle (29.82,16.54);
\begin{scope}[opacity=0.10]
  \foreach \sx/\sw in {0.75/7.05, 8.4/6.6, 15.6/6.9, 23.4/5.79}{
    \fill[black, rounded corners=12pt] (\sx+0.08,15.99-0.13) rectangle (\sx+\sw+0.08,15.99-12.64-0.13);
  }
\end{scope}
\node[panel, draw=borderblue,   fill=white, minimum width=7.05cm, minimum height=12.64cm, anchor=north west] (P1) at (0.75,15.99) {};
\node[panel, draw=bordergreen,  fill=white, minimum width=6.6cm, minimum height=12.64cm, anchor=north west] (P2) at (8.4,15.99) {};
\node[panel, draw=borderteal,   fill=white, minimum width=6.9cm, minimum height=12.64cm, anchor=north west] (P3) at (15.6,15.99) {};
\node[panel, draw=borderpurple, fill=white, minimum width=5.79cm, minimum height=12.64cm, anchor=north west] (P4) at (23.4,15.99) {};
\fill[panelblue!4, rounded corners=12pt] (0.82,15.92) rectangle (7.73,11.99);
\fill[panelgreen!4, rounded corners=12pt] (8.47,15.92) rectangle (14.93,11.99);
\fill[panelteal!4, rounded corners=12pt] (15.67,15.92) rectangle (22.43,11.99);
\fill[panelpurple!4, rounded corners=12pt] (23.47,15.92) rectangle (29.12,11.99);
\foreach \bx/\bc in {1.27/panelblue, 8.92/panelgreen, 16.12/panelteal, 23.92/panelpurple}{
  \draw[\bc, opacity=0.25, line width=1pt] (\bx,15.46) circle (0.42);
}
\node[badge, fill=panelblue] at (1.27,15.46) {1};
\node[anchor=west, font=\sffamily\bfseries, text=panelblue, align=left, scale=1.08] at (1.73,15.46) {PUBLIC DISCOURSE};
\node[badge, fill=panelgreen] at (8.92,15.46) {2};
\node[anchor=west, font=\sffamily\bfseries, text=panelgreen, align=left, scale=1.08] at (9.38,15.46) {BESPOKE INSTRUMENT};
\node[badge, fill=panelteal] at (16.12,15.46) {3};
\node[anchor=west, font=\sffamily\bfseries, text=panelteal, align=left, scale=1.05] at (16.58,15.46) {ACTIVATION \& MEDIATION};
\node[badge, fill=panelpurple] at (23.92,15.46) {4};
\node[anchor=west, font=\sffamily\bfseries, text=panelpurple, align=left, scale=1.0] at (24.38,15.46) {RESEARCH OUTPUT};
\draw[panelblue,  line width=1.6pt, line cap=round] (1.27,14.62)--(4.6,14.62);
\draw[panelgreen, line width=1.6pt, line cap=round] (8.92,14.62)--(12.2,14.62);
\draw[panelteal,  line width=1.6pt, line cap=round] (16.12,14.62)--(20.2,14.62);
\draw[panelpurple,line width=1.6pt, line cap=round] (23.92,14.62)--(27.0,14.62);
\begin{scope}
  \clip[rounded corners=10pt] (1.05,8.55) rectangle (7.10,13.05);
  \shade[top color=blue!7, bottom color=panelblue!16] (1.05,8.55) rectangle (7.10,13.05);
  \rainfield{(4.07,11.55)}{1.5}
\end{scope}
\draw[borderblue, line width=0.8pt, rounded corners=10pt] (1.05,8.55) rectangle (7.10,13.05);
\stormcloud{(2.55,12.35)}{1.18}
\speechbubbles{(5.20,11.95)}{1.30}
\floodhouse{(2.55,10.20)}{1.42}{-0.34}
\floodhouse{(4.55,10.02)}{1.10}{-0.30}
\treei{(6.05,10.12)}{1.30}
\floodwaterB{(4.07,9.50)}{1.88}
\draw[borderblue, line width=0.7pt] (1.15,6.35)--(7.40,6.35);
\database{(1.87,5.10)}{0.58}{dbBlight}{dbBdark}
\node[anchor=west, font=\sffamily\bfseries, text=bodytext, scale=0.92] at (2.82,5.65) {Meta Content Library};
\node[anchor=west, font=\sffamily, text=bodytext, align=left, scale=0.88] at (2.82,4.90) {Public flood-related discourse\\[4pt] 2012--2025};
\node[font=\sffamily\itshape, text=bodytext, scale=1.05] at (11.70,14.29) {Bespoke instrument};
\draw[panelgreen, line width=0.9pt] (11.00,13.84)--(12.40,13.84);
\node[dompill=bordergreen] at (12.20,12.39) {};
\heartbadge{(10.20,12.39)}{beLight}{beDark}
\node[anchor=west, font=\sffamily\bfseries\small, text=beDark, scale=1.2] at (11.05,12.71) {BE$-$};
\node[anchor=west, font=\sffamily, text=bodytext, align=left, scale=0.88] at (11.05,12.17) {\\[4pt]Affective--cognitive\\ distress appraisal}; 
\node[dompill=borderteal] at (12.20,10.49) {};
\personbadge{(10.20,10.49)}{doLight}{doDark}
\node[anchor=west, font=\sffamily\bfseries\small, text=doDark, scale=1.2] at (11.05,10.81) {DO$-$};
\node[anchor=west, font=\sffamily, text=bodytext, align=left, scale=0.88] at (11.05,10.27) {\\[4pt]Diminished\\ efficacy \& agency}; 
\node[dompill=borderpurple] at (12.20,8.59) {};
\groupbadge{(10.20,8.59)}{cnLight}{cnDark}
\node[anchor=west, font=\sffamily\bfseries\small, text=cnDark, scale=1.2] at (11.05,8.91) {CONNECT$+$};
\node[anchor=west, font=\sffamily, text=bodytext, align=left,scale=0.88] at (11.05,8.31) {\\[3pt]Relational--institutional\\ connectedness}; 
\draw[panelgreen, line width=1.2pt, line cap=round] (9.00,12.39)--(9.00,8.59);
\foreach \by in {12.39,10.49,8.59}{
  \draw[panelgreen, line width=1.2pt, line cap=round] (9.00,\by)--(9.52,\by);
  \fill[white] (9.52,\by) circle (0.075);
  \fill[panelgreen] (9.52,\by) circle (0.055);
}
\draw[panelgreen, line width=1.2pt, line cap=round] (8.72,10.49)--(9.30,10.49);
\fill[white] (9.00,10.49) circle (0.115);
\shade[inner color=panelgreen!70, outer color=panelgreen] (9.00,10.49) circle (0.085);
\node[panel, draw=bordergreen, fill=green!6, minimum width=6.0cm, minimum height=1.55cm] (RB) at (11.70,6.35) {};
\brainnet{(9.95,6.35)}{0.55}
\node[anchor=west, font=\sffamily\bfseries, text=panelgreen, scale=0.95] at (10.65,6.62) {Wellbeing reasoning};
\node[anchor=west, font=\sffamily, text=bodytext, scale=0.88] at (10.65,6.12) {(at population scale)};
\greycloud{(19.05,13.65)}{0.92}
\node[font=\sffamily\bfseries, text=bodytext, scale=0.95] at (19.05,12.62) {Flood stressors};
\draw[panelpurple!75, line width=1.1pt, line cap=round, line join=round]
  (19.05,12.36)--(19.05,12.14);
\draw[panelpurple!75, line width=1.1pt, line cap=round, line join=round]
  (16.65,12.14)--(21.45,12.14);
\foreach \mx in {16.65,19.05,21.45}{
  \draw[-{Triangle[length=2.3mm,width=2.6mm]}, panelpurple!75, line width=1.1pt, line cap=round] (\mx,12.14)--(\mx,11.84);
}
\fill[white] (19.05,12.14) circle (0.07);\fill[panelpurple!75] (19.05,12.14) circle (0.05);
\node[pathbox, minimum width=6.2cm, minimum height=2.55cm, anchor=north] at (19.05,11.78) {};
\node[badge, fill=panelblue, scale=0.66] at (16.20,11.45) {1};
\node[font=\sffamily\bfseries, text=panelblue, scale=0.92] at (19.35,11.45) {Stressor--domain activation};
\stormcloud{(16.70,10.78)}{0.44}\node[font=\sffamily,scale=0.58,text=bodytext,align=center] at (16.70,10.18){Flood\\ stressors};
\heartmini{(18.30,10.73)}{beLight}{beDark}\node[font=\sffamily\bfseries,scale=0.55,text=bodytext] at (18.30,10.18){BE$-$};
\personmini{(19.70,10.73)}{doLight}{doDark}\node[font=\sffamily\bfseries,scale=0.55,text=bodytext] at (19.70,10.18){DO$-$};
\groupmini{(21.30,10.73)}{cnLight}{cnDark}\node[font=\sffamily\bfseries,scale=0.46,text=bodytext] at (21.30,10.18){CONNECT$+$};
\draw[-{Stealth[length=2.0mm,width=2.2mm,round]}, panelteal!90, line width=1.4pt, line cap=round] (17.25,10.73)--(17.80,10.73);
\draw[-{Stealth[length=2.0mm,width=2.2mm,round]}, panelteal!90, line width=1.4pt, line cap=round] (18.80,10.73)--(19.20,10.73);
\draw[-{Stealth[length=2.0mm,width=2.2mm,round]}, panelteal!90, line width=1.4pt, line cap=round] (20.20,10.73)--(20.68,10.73);
\node[pathbox, minimum width=6.2cm, minimum height=2.5cm, anchor=north] at (19.05,8.94) {};
\node[badge, fill=panelteal, scale=0.66] at (16.20,8.61) {2};
\node[font=\sffamily\bfseries, text=panelteal, scale=0.92] at (19.35,8.61) {Appraisal--efficacy mediation};
\stormcloud{(16.85,7.94)}{0.44}\node[font=\sffamily,scale=0.58,text=bodytext,align=center] at (16.85,7.34){Flood\\ stressors};
\personmini{(19.05,7.89)}{doLight}{doDark}\node[font=\sffamily\bfseries,scale=0.55,text=bodytext] at (19.05,7.34){DO$-$};
\heartmini{(21.15,7.89)}{beLight}{beDark}\node[font=\sffamily\bfseries,scale=0.55,text=bodytext] at (21.15,7.34){BE$-$};
\draw[-{Stealth[length=2.0mm,width=2.2mm,round]}, panelteal!90, line width=1.4pt, line cap=round] (17.40,7.89)--(18.50,7.89);
\draw[-{Stealth[length=2.0mm,width=2.2mm,round]}, panelteal!90, line width=1.4pt, line cap=round] (19.60,7.89)--(20.60,7.89);
\node[pathbox, minimum width=6.2cm, minimum height=2.5cm, anchor=north] at (19.05,5.94) {};
\node[badge, fill=panelpurple, scale=0.66] at (16.20,5.61) {3};
\node[font=\sffamily\bfseries, text=panelpurple, scale=0.92] at (19.35,5.61) {Social--buffering mediation};
\stormcloud{(16.70,4.94)}{0.44}\node[font=\sffamily,scale=0.58,text=bodytext,align=center] at (16.70,4.34){Flood\\ stressors};
\groupmini{(18.30,4.89)}{cnLight}{cnDark}\node[font=\sffamily\bfseries,scale=0.46,text=bodytext] at (18.30,4.34){CONNECT$+$};
\personmini{(19.70,4.89)}{doLight}{doDark}\node[font=\sffamily\bfseries,scale=0.55,text=bodytext] at (19.70,4.34){DO$-$};
\heartmini{(21.30,4.89)}{beLight}{beDark}\node[font=\sffamily\bfseries,scale=0.55,text=bodytext] at (21.30,4.34){BE$-$};
\draw[-{Stealth[length=2.0mm,width=2.2mm,round]}, panelteal!90, line width=1.4pt, line cap=round] (17.25,4.89)--(17.68,4.89);
\draw[-{Stealth[length=2.0mm,width=2.2mm,round]}, panelteal!90, line width=1.4pt, line cap=round] (18.92,4.89)--(19.20,4.89);
\draw[-{Stealth[length=2.0mm,width=2.2mm,round]}, panelteal!90, line width=1.4pt, line cap=round] (20.20,4.89)--(20.80,4.89);
\database{(25.35,11.55)}{1.2}{dbPlight}{dbPdark}
\node[panel, draw=borderpurple, fill=white, minimum width=2.7cm, minimum height=2.2cm] at (26.70,10.05) {};
\charticon{(26.70,10.05)}{1.2}
\draw[borderpurple, line width=0.7pt] (24.00,6.95)--(28.60,6.95);
\node[font=\sffamily\bfseries, text=panelpurple, align=center, scale=1.16] at (26.30,5.95) {Activation \& mediation\\mechanisms,\\ multi-domain dataset};
\node[font=\sffamily, text=bodytext, align=center, scale=0.9] at (26.30,4.55) {Flood wellbeing in Ireland,\\ 2012--2025};
\foreach \ax/\bx/\colmid/\col in {7.62/8.58/arrowblue!60/arrowblue, 14.82/15.78/arrowgreen!60/arrowgreen, 22.47/23.43/arrowblue!60/arrowblue}{
  \fill[\col, opacity=0.09] ({(\ax+\bx)/2},10.49) ellipse (0.66 and 0.38);
  \fill[\col, opacity=0.07] ({(\ax+\bx)/2},10.49) ellipse (0.44 and 0.25);
  \draw[-{Triangle[length=3.9mm,width=5.1mm]}, line width=4.2pt, line cap=round, \colmid] (\ax,10.49)--(\bx,10.49);
  \draw[-{Triangle[length=3.4mm,width=4.4mm]}, line width=2.4pt, line cap=round, \col] (\ax,10.49)--(\bx,10.49);
}
\node[panel, draw=borderpurple, fill=white, minimum width=28.44cm, minimum height=2.0cm, anchor=center] (SUM) at (14.97,2.01) {};
\node[text width=26.44cm, font=\sffamily, text=bodytext, align=left, scale=1.0] at (14.97,2.01) {This study analyses \textbf{\textcolor{panelblue}{flood-related wellbeing in Ireland}} using large-scale public discourse, develops a bespoke three-domain wellbeing instrument, reveals \textbf{\textcolor{panelteal}{stressor activation and resilience mediation pathways}}, and delivers a \textbf{\textcolor{panelpurple}{multi-indicator, multi-domain temporal data}} for 2012--2025.};
\end{tikzpicture}
\endgroup
}
\end{center}
\end{graphicalabstract}

\begin{highlights}

     \item \textbf{A bespoke flood-wellbeing instrument for unobtrusive discourse analysis}: we develop a theory-informed instrument for large-scale analysis of flood-related public discourse, characterising flood wellbeing across \emph{affective--cognitive distress appraisal}, \emph{diminished perceived efficacy and agency}, and \emph{perceived relational--institutional connectedness}.

    \item \textbf{Flood-related social--ecological resilience activation and mediation}: using large-scale unobtrusive discourse from Meta Content Library, we show that \emph{flood stressors} activate \emph{affective--cognitive distress appraisal}, \emph{diminished perceived efficacy and agency}, and \emph{perceived relational--institutional connectedness}, with \emph{appraisal--efficacy} and \emph{social-buffering} mediation pathways surfacing in Irish flood-related discourse.

    \item \textbf{A multi-indicator, multi-domain temporal dataset for regional flood-wellbeing research}: we create a secondary temporal dataset of flood-related wellbeing in Ireland for 2012--2025, organised across multiple indicators and domains to support longitudinal research on wellbeing dynamics, disaster experience, climate adaptation, social--ecological resilience, and public responses to flooding.
    
\end{highlights}

\begin{keywords}
Subjective wellbeing \sep
Flood vulnerability \sep
Flood risk management \sep
Flood-related wellbeing \sep
Irish Flood discourse \sep
Unobtrusive discourse analysis \sep
Climate adaptation
\end{keywords}

\maketitle

\section{Introduction}



Flooding is a climate-related environmental hazard with significant social and psychological consequences for exposed individuals, shaping threat appraisal \citep{LazarusFolkman1984_StressAppraisal}, perceived coping capacity \citep{Bandura1977_SelfEfficacy, Bandura1997_SelfEfficacyControl}, access to social support \citep{CohenWills1985_SupportBuffering,KaniastyNorris1993_SupportDeterioration}, and confidence in institutional response \citep{Rousseau1998_Trust}. Flood-related wellbeing refers to the ways in which flood exposure shapes individual affective appraisal, perceived agency, daily functioning, access to social support, and confidence in institutional response. Beyond the direct material and economic consequences of flooding, understanding the pathway structures and activation and mediation mechanisms across these domains is therefore critical for assessing the mental-health and wellbeing impacts of flood exposure and for informing stakeholder decision-making in climate adaptation policy.

Disaster research has commonly examined these psychosocial dimensions through post-event surveys and interviews that elicit self-reported distress, resource disruption, coping efficacy, and social support following exposure \citep{LazarusFolkman1984_StressAppraisal, Hobfoll1989_COR, CohenWills1985_SupportBuffering, Bandura1977_SelfEfficacy, Norris2002_60kVictims}. However, surveys and interviews, as structured and researcher-elicited modes of inquiry, are costly and time-intensive, which constrains their feasibility for large-scale population-level research. The samples generated by such designs are therefore often relatively small and potentially subject to selection bias, limiting their capacity to capture the wider non-elicited public discourse through which flood experiences are articulated during and after events.

In response to the need for population-level analysis of flood-related wellbeing and its psychosocial consequences, and to address the methodological constraints of structured researcher-elicited approaches, this paper develops an unobtrusive discourse-based method for analysing flood exposure through naturally occurring public discourse. We analyse large-scale, non-elicited discourse from Meta Content Library (MCL) \citep{Meta_MCL_v6}, comprising approximately 24{,}000$+$ flood-related public posts in Ireland spanning the period 2012--2025. Unlike structured survey or interview responses, these materials consist of unstructured discourse produced organically during and after flood events, capturing how individuals publicly articulate flood threat, disruption, coping, support, and recovery as events unfold. To enable population-level analysis of these large-scale unstructured discourse records, we develop a flood-wellbeing reasoning pipeline with large language models (LLMs) \citep{Brown2020_GPT3,Ouyang2022_InstructGPT} to extract multi-indicator, multi-domain wellbeing measures from public discourse at scale.

Social--ecological resilience refers to the capacity of individuals and communities to absorb, adapt to, and recover from environmental stressors through interconnected psychological, social, and institutional processes \citep{Hobfoll1989_COR, Norris2008_CommunityResilience}. In disaster settings, resilience is understood as arising from the interaction between hazard exposure, affective appraisal and coping efficacy, social and institutional resources, and subsequent wellbeing and recovery trajectories \citep{LazarusFolkman1984_StressAppraisal, Bandura1977_SelfEfficacy, CohenWills1985_SupportBuffering, Norris2008_CommunityResilience}. Our study reveals stressor activation and social--ecological resilience mediation pathways in Irish flood discourse. First, flood exposure directly activates flood-related wellbeing domains, including affective appraisal, perceived coping efficacy, and relational--institutional connectedness. Second, the appraisal--efficacy mediation pathway further links flood exposure to threat appraisal, psychological distress, and diminished perceived efficacy and agency, while reduced efficacy may in turn reinforce negative appraisal and distress. 

Similar appraisal--efficacy mediation dynamics have previously been reported in disaster research, including studies of survivors of \emph{Hurricane Andrew} (August, 1992) \citep{Benight1999b_HurricaneAndrewModel} and the \emph{Mount St.~Helens} volcanic eruption (May, 1980) \citep{Murphy1988_MtStHelens}, where coping self-efficacy was found to mediate psychological adjustment under disaster exposure \citep{Bandura1977_SelfEfficacy, LazarusFolkman1984_StressAppraisal}. The social-buffering mediation pathway links social and institutional protective resources to perceived efficacy and subsequent appraisal of flood-related adversity \citep{CohenWills1985_SupportBuffering, KaniastyNorris1995_AltruisticCommunity, Hobfoll1989_COR}. Comparable social-buffering dynamics have been reported in stress and disaster research, including the general stress-buffering model of social support \citep{CohenWills1985_SupportBuffering}, post-disaster support mobilisation and the altruistic-community phase following \emph{Hurricane Hugo} (September, 1989) \citep{KaniastyNorris1995_AltruisticCommunity}, and resource-based accounts of stress and recovery in \emph{Conservation of Resources} theory \citep{Hobfoll1989_COR}. Together, these studies support the interpretation that relational and protective resources can strengthen perceived coping capacity and reduce psychological burden during post-disaster adjustment.


Although prior disaster and resilience research has documented appraisal--efficacy and social-buffering relationships across a range of disaster contexts, flood-wellbeing research has given comparatively limited attention to their activation and mediation dynamics under recurrent flood exposure. In particular, little research has examined how flood stressors directly activate multiple wellbeing domains, or how affective appraisal, perceived efficacy, and social protection operate as social--ecological resilience mediation pathways in large-scale public discourse over time. Our findings show that these activation and mediation pathways surface within Irish flood-related discourse and can be examined longitudinally through discourse-based flood-wellbeing indicators. Our contributions are as follows:
\begin{itemize}
    \item \textbf{A bespoke flood-wellbeing instrument.} We develop a theory-informed bespoke flood-wellbeing instrument for unobtrusive discourse-based wellbeing study, organised into three overarching domains: affective--cognitive distress appraisal, diminished perceived efficacy and agency, and perceived relational--institutional connectedness.

    \item \textbf{Activation and mediation mechanisms.} We identify activation and social--ecological resilience mediation pathways in Irish flood-related discourse, showing that flood stressors directly activate multiple wellbeing domains and that appraisal--efficacy and social-buffering mediation pathways structure flood-related wellbeing.

    \item \textbf{Regional flood-wellbeing data.} We construct a secondary multi-indicator, multi-domain temporal dataset of flood-related wellbeing in Ireland spanning 2012--2025, supporting longitudinal research on wellbeing, adaptation, resilience, and recurrent flood exposure.
\end{itemize}

\section{Background}

\paragraph{\textbf{Appraisal--efficacy mediation theory.}}
Research in disaster psychology has consistently shown that psychological adjustment under disaster exposure is shaped by the interaction between cognitive appraisal and perceived coping efficacy. Transactional theories of stress conceptualise disasters as stressors that activate processes of threat appraisal, uncertainty assessment, and perceived controllability \citep{LazarusFolkman1984_StressAppraisal}. Self-efficacy theory further proposes that perceived capability to manage adversity strongly influences emotional regulation, coping behaviour, and post-disaster adjustment \citep{Bandura1977_SelfEfficacy, Bandura1997_SelfEfficacyControl}. Within disaster contexts, appraisal and efficacy therefore form an interconnected pathway: heightened threat appraisal and psychological distress are often associated with diminished perceived efficacy and agency, while reduced efficacy may in turn reinforce negative appraisal and emotional burden. Such appraisal--efficacy dynamics have been documented across multiple disaster settings, including studies of survivors of \emph{Hurricane Andrew} and the \emph{Mount St.~Helens} volcanic eruption, where coping self-efficacy mediated psychological adjustment following disaster exposure \citep{Benight1999b_HurricaneAndrewModel, Murphy1988_MtStHelens}. Although these relationships are well established within disaster psychology, comparatively little flood-related wellbeing research has examined how appraisal and efficacy pathways are jointly activated and sustained under recurrent flood exposure.

\paragraph{\textbf{Social--buffering mediation theory.}}
A second major line of disaster research concerns the protective role of social and institutional resources during periods of environmental disruption. Stress-buffering theory proposes that social support mitigates the psychological consequences of stress by providing emotional reassurance, informational guidance, and practical assistance during adversity \citep{CohenWills1985_SupportBuffering}. Related work in disaster resilience further shows that collective solidarity, support mobilisation, and institutional trust are important determinants of post-disaster adaptation and recovery \citep{KaniastyNorris1993_SupportDeterioration, KaniastyNorris1995_AltruisticCommunity}. Resource-based theories of resilience similarly emphasise that access to social, material, and institutional resources shapes the capacity of individuals and communities to absorb and recover from disaster-related disruption \citep{Hobfoll1989_COR}. Prior research has therefore documented strong associations between social protection, perceived efficacy, and psychological adjustment following disasters. However, flood-related wellbeing research has given comparatively limited attention to how social-buffering mechanisms co-occur with appraisal--efficacy dynamics under recurrent flood exposure.

\paragraph{\textbf{Positioning in flood-related wellbeing research.}}
Wellbeing in disaster research is commonly understood as a multidimensional condition encompassing affective functioning, perceived agency, social connectedness, and the capacity to maintain everyday functioning under environmental stress \citep{LazarusFolkman1984_StressAppraisal, Hobfoll1989_COR, Bandura1977_SelfEfficacy, CohenWills1985_SupportBuffering}. Flood-related wellbeing therefore concerns the ways in which flood exposure shapes affective appraisal, perceived agency, daily functioning, social support, and confidence in institutional response. Empirical flood-mental-health research has documented these consequences across a range of flood contexts, including the 1999 Mexico flood \citep{Fernandez2015_FloodMentalHealth}, the 2002 Dresden flood in Germany \citep{Otto2006_FloodJustWorld}, the 2007 summer floods in England \citep{Stanke2012_FloodingMentalHealth}, flood events in Hunan, China \citep{Fernandez2015_FloodMentalHealth}, Storm Xynthia in France \citep{Fernandez2015_FloodMentalHealth}, and the 2021 European floods \citep{Li2023_FloodSentimentEurope}. This literature shows that flood exposure is associated with psychological distress, anxiety, post-traumatic stress symptoms, disruption to daily life, secondary stressors, and longer-term psychosocial burden \citep{Stanke2012_FloodingMentalHealth, Fernandez2015_FloodMentalHealth}. 

However, existing flood-wellbeing research has primarily been organised around empirical case studies using post-event surveys, interviews, clinical reports, assessment instruments, or sentiment-based computational analysis. These approaches provide important evidence on flood-related mental-health outcomes, but they have given comparatively limited attention to activation and mediation mechanisms linking flood stressors, affective appraisal, perceived efficacy, functional disruption, and social protection as an integrated pathway structure. Survey- and interview-based designs also remain costly, time-intensive, and often limited in population-scale coverage, while sentiment-based social-media approaches are scalable but typically reduce public response to a single polarity dimension \citep{BryanSmith2023_FloodSentiment, Li2023_FloodSentimentEurope}. These limitations motivate an unobtrusive discourse-based approach that can study flood-related wellbeing at scale while preserving a multidimensional construct structure aligned with disaster, coping, and resilience theory.

\section{Development of bespoke instrument}

\begin{table}[pos=b]

\centering

\caption{Definition of nine wellbeing indicators, which are organised into three polarised BE/DO/CONNECT
domains. The sign $+$/$-$ marks the domain polarity.}\label{tab:indicators}

\small

\resizebox{0.9\linewidth}{!}{
\begin{tabular}{p{2.0cm}  p{4.3cm} p{9cm}}
\toprule
\textbf{Domain} & \textbf{Indicator} & \textbf{Construct definition} \\
\midrule

\multirow{3}{*}{\parbox{1\linewidth}{\vspace{1cm} \textbf{BE$-$} }}
&  \cellcolor{blue!08}\textbf{I1: Stress}
& \cellcolor{blue!08}Subjective appraisal of flood-related demands as exceeding coping capacity, resulting in overwhelm and tension. \\

& \cellcolor{teal!08}\textbf{I2: Anxiety}
& \cellcolor{teal!08}Persistent worry, fear, dread, and hypervigilance regarding flood-related threats and uncertainty. \\

& \cellcolor{pink!08}\textbf{I3: Hopelessness}
& \cellcolor{pink!08}Negative expectations regarding future outcomes, including diminished agency thinking and pathways thinking. \\
\midrule
\multirow{3}{*}{\parbox{1\linewidth}{\vspace{1cm} \textbf{DO$-$} }}

& \cellcolor{blue!08}\textbf{I4: Low efficacy}
& \cellcolor{blue!08}Perceived inability to mobilise cognitive, behavioural, and social resources to manage flood-related adversity. \\

& \cellcolor{teal!08}\textbf{I5: Low daily functioning}
& \cellcolor{teal!08}Disruption of housing stability, occupational and domestic role performance, and routine daily functioning under flood exposure. \\

& \cellcolor{pink!08}\textbf{I6: Limited access to resources}
& \cellcolor{pink!08}Lack of, or impeded access to, material, economic, and service resources required for adaptation and recovery. \\
\midrule
\multirow{3}{*}{\parbox{1\linewidth}{\vspace{1cm} \textbf{CONNECT$+$} }}

& \cellcolor{blue!08}\textbf{I7: Connectedness}
& \cellcolor{blue!08}Sense of social belonging, relational embeddedness, and integration within meaningful interpersonal networks. \\

& \cellcolor{teal!08}\textbf{I8: Protective support}
& \cellcolor{teal!08}Availability or receipt of instrumental, emotional, and informational support from family, friends, neighbours, and the wider community. \\

& \cellcolor{pink!08}\textbf{I9: Institutional trust}
& \cellcolor{pink!08}Confidence in the reliability, competence, and responsiveness of institutions and emergency services. \\
\bottomrule
\end{tabular}
}


\vspace{2em}

\caption{Impact scale rubrics used for scoring.}\label{tab:scales}

\resizebox{0.9\linewidth}{!}{
\begin{tabular}{c l p{10cm}}

\toprule
\textbf{Scale} & \textbf{Descriptor} & \textbf{Rubric note} \\
\midrule
0 & \textbf{No evidence}
 & The material contains no evidence relevant to this indicator
   in the specified temporal phase. The material producer does not \emph{mention}, \emph{imply}, or \emph{allude} to any experience related to this indicator
   during this phase. \\[4pt]
1 & \textbf{Low impact}
 & The material contains a mild or brief mention relevant to this
   indicator. The reference is incidental rather than central to the narrative and account of the content producer --- for example, a passing remark or a single word or phrase that touches on the indicator without elaboration or emotional weight. \\[4pt]
2 & \textbf{Moderate impact}
 & The material contains clear and elaborated evidence relevant to
   this indicator. The content producer describes the experience in some
   detail, conveys discernible emotional or functional consequence,
   and the indicator is recognisably present as a meaningful element
   of the account of the content producer, though not necessarily the dominant theme. \\[4pt]
3 & \textbf{Severe impact}
 & The material contains intense, sustained, or dominant expression
   relevant to this indicator. The experience is a central theme of the account of the content producer, conveyed with strong emotional weight, detailed description, or repeated reference. The indicator represents a major source of distress, disruption, or concern in the specified temporal phase. \\
\bottomrule
\end{tabular}
}

\end{table}
\begin{table}[pos=b]

\centering
\caption{Assessment instrument. Evidence-based impact scale per indicator on the 0~+~1--3 scale (see Table~\ref{tab:scales}). Indicators are defined in Table~\ref{tab:indicators}.}
\label{tab:instrument_nophase}

\begin{tabular}{l p{8cm} l}
\toprule
\textbf{Indicator} & \textbf{Question} & \textbf{Score} \\
\midrule

\textbf{I1: Stress}
 & Based on the material, to what extent does the content producer express or describe stress, emotional overwhelm, or an inability to cope with demands related to flood experience?
 & 0\enspace 1\enspace 2\enspace 3 \\[6pt]

\textbf{I2: Anxiety}
 & Based on the material, to what extent does the content producer express or describe worry, fear, dread, or hypervigilance related to flood-related threats?
 & 0\enspace 1\enspace 2\enspace 3 \\[6pt]

\textbf{I3: Hopelessness}
 & Based on the material, to what extent does the content producer express or describe hopelessness, pessimism, or loss of positive expectation regarding their situation and future outcomes?
 & 0\enspace 1\enspace 2\enspace 3 \\[6pt]

\textbf{I4: Low efficacy}
 & Based on the material, to what extent does the content producer express or describe a perceived inability or lack of capability to cope with flood-related adversity?
 & 0\enspace 1\enspace 2\enspace 3 \\[6pt]

\textbf{I5: Low daily functioning}
 & Based on the material, to what extent does the content producer describe disruption of housing stability, occupational and domestic routines, or physical activity under flood-related demands?
 & 0\enspace 1\enspace 2\enspace 3 \\[6pt]

\textbf{I6: Limited access to resources}
 & Based on the material, to what extent does the content producer describe lack of, or impeded, access to essential material, economic, and service resources---including shelter, employment, income, education, and healthcare---under flood-related conditions?
 & 0\enspace 1\enspace 2\enspace 3 \\[6pt]

\textbf{I7: Connectedness}
 & Based on the material, to what extent does the content producer express or describe a sense of social belonging, relational embeddedness, or integration within meaningful interpersonal networks in relation to flood experience?
 & 0\enspace 1\enspace 2\enspace 3 \\[6pt]

\textbf{I8: Protective support}
 & Based on the material, to what extent does the content producer describe the availability or receipt of instrumental, emotional, or informational help from family, friends, neighbours, or the wider community in relation to flood experience?
 & 0\enspace 1\enspace 2\enspace 3 \\[6pt]

\textbf{I9: Institutional trust}
 & Based on the material, to what extent does the content producer express or describe confidence in the reliability, competence, and responsiveness of authorities, institutions, and emergency services with respect to flood response and recovery?
 & 0\enspace 1\enspace 2\enspace 3 \\

\bottomrule
\end{tabular}

\end{table}

\begin{figure}
    \centering
    
    \resizebox{0.86\linewidth}{!}{\begingroup
\definecolor{cream}{HTML}{FBF6EC}
\definecolor{inkblue}{HTML}{1F4E6B}
\definecolor{floodfill}{HTML}{E7EFF5}
\definecolor{floodring}{HTML}{6E9AB8}
\definecolor{befill}{HTML}{EFE6EC}
\definecolor{bering}{HTML}{9A7B93}
\definecolor{connfill}{HTML}{E6EEDF}
\definecolor{connring}{HTML}{6E8C5A}
\definecolor{dofill}{HTML}{F7EBD6}
\definecolor{doring}{HTML}{C79A55}
\definecolor{mauvearr}{HTML}{8E5E86}
\definecolor{greenlab}{HTML}{4E7A3A}
\definecolor{amberlab}{HTML}{C0892F}
\definecolor{legendbg}{HTML}{F2ECE0}
\def\iconcloud#1{%
\begin{scope}[shift={#1}, scale=0.40, draw=floodring, line width=1.1pt, line cap=round, line join=round]
  \draw[fill=floodring!12] (-0.9,0.35) .. controls (-1.2,0.35) and (-1.2,0.95) .. (-0.7,0.95)
    .. controls (-0.6,1.3) and (0.1,1.35) .. (0.35,1.05)
    .. controls (0.75,1.2) and (1.15,0.8) .. (0.8,0.45)
    .. controls (1.0,0.1) and (0.55,-0.15) .. (0.25,0.1)
    .. controls (0.0,-0.05) and (-0.7,-0.05) .. (-0.9,0.35) -- cycle;
  \foreach \x in {-0.6,-0.2,0.2,0.6} { \draw (\x,-0.05) -- (\x,-0.5); }
  \draw (-0.9,-0.8) .. controls (-0.6,-0.6) and (-0.3,-1.0) .. (0,-0.8)
        .. controls (0.3,-0.6) and (0.6,-1.0) .. (0.9,-0.8);
\end{scope}}
\def\iconbrain#1{%
\begin{scope}[shift={#1}, scale=0.48, draw=bering, line width=1.0pt, line cap=round, line join=round]
  \draw[fill=bering!12] (0,0.7)
    .. controls (-0.55,0.95) and (-1.0,0.55) .. (-0.85,0.1)
    .. controls (-1.1,-0.25) and (-0.7,-0.6) .. (-0.35,-0.5)
    .. controls (-0.2,-0.75) and (0.2,-0.75) .. (0.35,-0.5)
    .. controls (0.7,-0.6) and (1.1,-0.25) .. (0.85,0.1)
    .. controls (1.0,0.55) and (0.55,0.95) .. (0,0.7) -- cycle;
  \draw (0,0.7) -- (0,-0.6);
  \draw (-0.3,0.4) .. controls (-0.1,0.3) .. (-0.3,0.1);
  \draw (0.3,0.4) .. controls (0.1,0.3) .. (0.3,0.1);
  \draw (-0.3,-0.15) .. controls (-0.1,-0.25) .. (-0.25,-0.4);
  \draw (0.3,-0.15) .. controls (0.1,-0.25) .. (0.25,-0.4);
\end{scope}}
\def\iconpeople#1{%
\begin{scope}[shift={#1}, scale=0.46, draw=connring, line width=1.0pt, line cap=round, line join=round]
  \draw[fill=connring!12] (-0.7,0.55) circle (0.22);
  \draw[fill=connring!12] (0.7,0.55) circle (0.22);
  \draw (-1.05,-0.15) .. controls (-1.0,0.25) and (-0.4,0.25) .. (-0.35,-0.15);
  \draw (0.35,-0.15) .. controls (0.4,0.25) and (1.0,0.25) .. (1.05,-0.15);
  \draw[fill=connring!12] (0,0.65) circle (0.28);
  \draw (-0.5,-0.35) .. controls (-0.45,0.2) and (0.45,0.2) .. (0.5,-0.35) -- cycle;
\end{scope}}
\def\icongear#1{%
\begin{scope}[shift={#1}, scale=0.46, draw=doring, line width=1.0pt, line cap=round, line join=round]
  \draw[fill=doring!12] (-0.55,0.55) circle (0.25);
  \draw (-0.95,-0.5) .. controls (-0.9,0.15) and (-0.2,0.15) .. (-0.15,-0.5) -- cycle;
  \def\R{0.42}\def\r{0.18}
  \draw[fill=doring!12] (0.55,0.15) circle (\r);
  \foreach \a in {0,45,...,315}{ \draw ($(0.55,0.15)+(\a:\r)$) -- ($(0.55,0.15)+(\a:\R)$); }
  \draw (0.55,0.15) circle (\R);
  \draw[fill=white] (0.55,0.15) circle (0.12);
\end{scope}}
\begin{tikzpicture}[>=Stealth, font=\sffamily,
  dcircle/.style={circle, line width=1pt, minimum size=4.0cm, inner sep=0pt, align=center}]
\fill[white] (-1.8,-7) rectangle (15.8,7.8);
\begin{scope}[opacity=0.35]
  \fill[floodring!18] (1,5) ellipse (2.7cm and 2.5cm);
  \fill[bering!18] (13,5) ellipse (2.7cm and 2.5cm);
  \fill[connring!16] (1,-1) ellipse (2.7cm and 2.5cm);
  \fill[doring!18] (13,-1) ellipse (2.7cm and 2.5cm);
\end{scope}
\node[dcircle, draw=floodring, fill=floodfill, text=inkblue] (S) at (1,5)
  {\\[14pt]\textbf{\large Flood}\\\textbf{\large Stressors}};
\node[dcircle, draw=bering, fill=befill, text=inkblue] (BE) at (13,5)
  {\\[14pt]\textbf{\large BE$-$}\\[1pt]\footnotesize (Affective--Cognitive\\ Distress)};
\node[dcircle, draw=connring, fill=connfill, text=greenlab] (CONN) at (1,-1)
  {\\[14pt]\textbf{\large CONNECT+}\\[1pt]\footnotesize (Relational--Institutional\\ Ecology of Recovery)};
\node[dcircle, draw=doring, fill=dofill, text=amberlab] (DO) at (13,-1)
  {\\[14pt]\textbf{\large DO$-$}\\[1pt]\footnotesize (Agentic \& Functional\\ Adaptive Capacity\\ Deficits)};
\iconcloud{($(S)+(0,0.95)$)}
\iconbrain{($(BE)+(0,1.05)$)}
\iconpeople{($(CONN)+(0,1.05)$)}
\icongear{($(DO)+(0,1.05)$)}
\draw[->, line width=1.4pt, draw=inkblue] (S) -- node[font=\itshape\large, text=inkblue, above] {Activate} (BE);
\draw[->, line width=1.4pt, draw=inkblue] (S) -- node[font=\itshape\large, text=inkblue, left] {Activate} (CONN);
\draw[->, line width=1.4pt, draw=inkblue] (S) -- node[font=\itshape\large, text=inkblue, above, sloped, pos=0.62] {Activate} (DO);
\draw[->, line width=1.2pt, draw=floodring, dashed, dash pattern=on 7pt off 5pt]
    (CONN) -- node[font=\itshape\large, text=floodring, below] {Mediate} (DO);
\draw[->, line width=1.1pt, draw=floodring, dashed, dash pattern=on 6pt off 4pt]
    ($(DO.north)+(-0.3,0)$) -- ($(BE.south)+(-0.3,0)$);
\draw[->, line width=1.1pt, draw=mauvearr, dotted, line cap=round]
    ($(DO.north)+(0.3,0)$) -- ($(BE.south)+(0.3,0)$)
    node[font=\itshape\large, text=mauvearr, midway, right=5pt] {Mediate};
\draw[->, line width=1.2pt, draw=connring, dash pattern=on 8pt off 3pt on 1pt off 3pt]
    (CONN) -- node[font=\itshape\large, text=connring, sloped, above, pos=0.30] {Mediate} (BE);
\begin{scope}[shift={(0,-5)}]
  \node[anchor=west, font=\bfseries\large, text=inkblue] at (0,0.9) {Direct activation pathways};
  \draw[->, line width=1.4pt, draw=inkblue] (0.2,-0.1) -- (1.6,-0.1);
  \node[anchor=west, font=\large, align=left] at (1.85,-0.1)
    {Flood Stressors directly activate\\ BE$-$, DO$-$, and CONNECT+.};
\end{scope}
\begin{scope}[shift={(9,-5)}]
  \node[anchor=west, font=\bfseries\large, text=inkblue] at (0,0.9) {Mediation pathways};
  \draw[->, line width=1.1pt, draw=floodring, dashed, dash pattern=on 6pt off 4pt] (0.2,0) -- (1.6,0);
  \node[anchor=west, font=\large] at (1.85,0)
    {Flood Stressors $\rightarrow$ \textcolor{greenlab}{\bfseries\itshape CONNECT+} $\rightarrow$ \textcolor{amberlab}{\bfseries\itshape DO$-$} $\rightarrow$ \textcolor{mauvearr}{\bfseries\itshape BE$-$}};
  \draw[->, line width=1.1pt, draw=mauvearr, dotted, line cap=round] (0.2,-0.8) -- (1.6,-0.8);
  \node[anchor=west, font=\large] at (1.85,-0.8)
    {Flood Stressors $\rightarrow$ \textcolor{amberlab}{\bfseries\itshape DO$-$} $\rightarrow$ \textcolor{mauvearr}{\bfseries\itshape BE$-$}};
  \draw[->, line width=1.2pt, draw=connring, dash pattern=on 8pt off 3pt on 1pt off 3pt] (0.2,-1.6) -- (1.6,-1.6);
  \node[anchor=west, font=\large] at (1.85,-1.6)
    {Flood Stressors $\rightarrow$ \textcolor{greenlab}{\bfseries\itshape CONNECT+} $\rightarrow$ \textcolor{mauvearr}{\bfseries\itshape BE$-$}};
\end{scope}
\begin{pgfonlayer}{background}
\end{pgfonlayer}
\end{tikzpicture}
\endgroup}
    
    \caption{\textbf{Activation and Mediation Hypotheses.} We hypothesise three direct activation pathways, in which \textbf{\emph{flood stressors}} activate the domains \textbf{BE$-$}, \textbf{DO$-$}, and \textbf{CONNECT$+$}. We further hypothesise two major mediation pathways: \textbf{\emph{flood stressors}} $\rightarrow$ \textbf{DO$-$} $\rightarrow$ \textbf{BE$-$}, and \textbf{\emph{flood stressors}} $\rightarrow$ \textbf{CONNECT$+$} $\rightarrow$ \textbf{DO$-$} $\rightarrow$ \textbf{BE$-$}. Although a mediation pathway, \textbf{\emph{flood stressors}} $\rightarrow$ \textbf{CONNECT$+$} $\rightarrow$ \textbf{BE$-$}, is also evident, the correlation appears weaker than the appraisal--efficacy pathway, \textbf{\emph{flood stressors}} $\rightarrow$ \textbf{DO$-$} $\rightarrow$ \textbf{BE$-$}, and the social-buffering pathway, \textbf{\emph{flood stressors}} $\rightarrow$ \textbf{CONNECT$+$} $\rightarrow$ \textbf{DO$-$} $\rightarrow$ \textbf{BE$-$}. This suggests that \textbf{CONNECT$+$} operates less as a direct suppressor of \textbf{BE$-$} than as a relational--institutional resource pathway that consolidates and strengthens perceived efficacy. In this interpretation, social support, protective mobilisation, and institutional confidence primarily mediate flood-related appraisal through \textbf{DO$-$}, by reducing efficacy deficits and strengthening perceived agency under flood exposure.}
    
    \label{fig:wellbeing_hypotheses}
\end{figure}
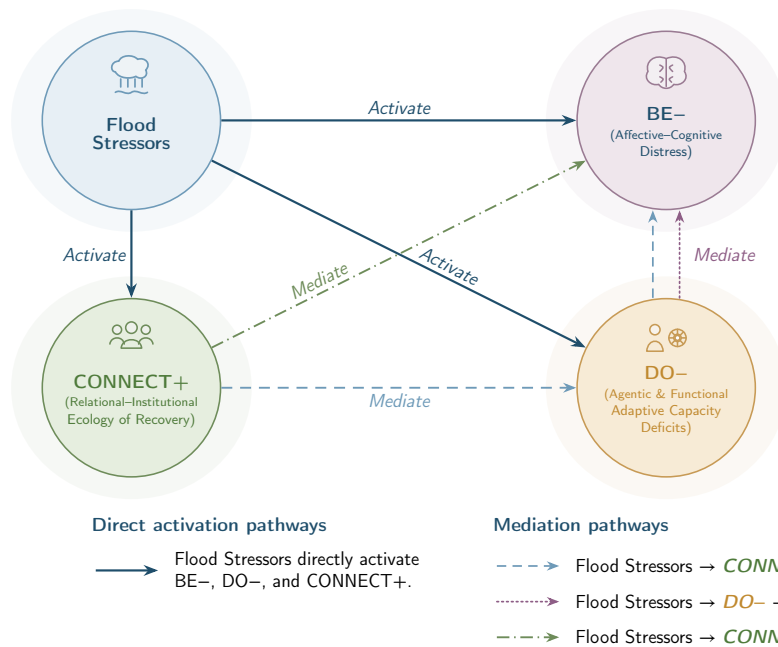

\paragraph{\textbf{Bespoke instrument, indicators, and domains.}}
To study flood-related wellbeing through unobtrusive public discourse, we develop a bespoke flood-wellbeing instrument designed for wellbeing reasoning with large language models on large-scale unstructured discourse data from Meta Content Library. Unlike conventional survey instruments administered through direct elicitation, the present instrument is designed for non-elicited public discourse, where experiences of flooding are embedded within naturally occurring language produced during and after flood events. The instrument operationalises flood-related wellbeing as a multidimensional construct organised into three theoretically grounded domains: BE$-$, DO$-$, and CONNECT$+$, where $+/-$ indicates the domain polarity. The tripartite structure is aligned with established theories of stress appraisal, self-efficacy, social buffering, and post-disaster resilience, and reflects the activation and mediation pathways examined in this study. As shown in Table~\ref{tab:indicators}, the flood-related wellbeing domains are organised as:
\begin{itemize}
    \item BE$-$ (\emph{affective--cognitive distress appraisal}) captures pathways of affective arousal, threat appraisal, and emotional dysregulation activated by flood exposure. The domain includes indicators of stress, anxiety, and hopelessness, reflecting negative-valence psychological states commonly associated with disaster exposure \citep{LazarusFolkman1984_StressAppraisal, Cohen1983_PSS, Norris2002_60kVictims, Snyder2002_HopeTheory}. 
    
    \item DO$-$ (\emph{diminished perceived efficacy and agency}) captures pathways of diminished coping efficacy, reduced personal agency, and impaired capacity to maintain everyday functioning under adversity. The domain includes indicators of efficacy deficits, functional disruption, and resource constraints, reflecting reductions in adaptive capacity commonly associated with disaster exposure and recovery challenges \citep{Bandura1977_SelfEfficacy, Benight1999a_HCSE, Hobfoll1989_COR, Norris2008_CommunityResilience}. 
    
    \item CONNECT$+$ (\emph{perceived relational--institutional connectedness}) captures pathways of social embeddedness, interpersonal support mobilisation, and institutional confidence through which the wider social ecology moderates disaster recovery. The domain includes indicators of connectedness, protective support, and institutional trust, reflecting relational and institutional protective factors associated with psychological adjustment following disasters \citep{Russell1996_UCLA, CohenWills1985_SupportBuffering, KaniastyNorris1993_SupportDeterioration, Rousseau1998_Trust}.
\end{itemize}

\paragraph{\textbf{Rating scales.}}
Unobtrusive discourse data do not provide access to respondents' pre-disaster baselines, clinical histories, or longitudinal psychological profiles. The instrument is therefore not intended to produce continuous psychometric severity estimates comparable to questionnaire-based wellbeing scales. Instead, we adopt an evidence-based ordinal rating framework designed for inference under conditions of partial and indirect behavioural evidence. As shown in Table~\ref{tab:scales}, each indicator is scored according to the strength of evidence identifiable within the discourse record, using ordinal categories corresponding to low, intermediate, and high evidential intensity. The scoring process therefore evaluates the degree to which theoretically defined wellbeing constructs are substantiated within discourse, rather than attempting direct clinical estimation of latent psychological states. This evidence-based categorical framework reduces inferential uncertainty in large-scale unstructured discourse analysis while preserving theoretically meaningful distinctions across multidimensional wellbeing domains.

Table~\ref{tab:indicators} defines the indicators and their associated wellbeing domains. Table~\ref{tab:scales} presents the evidence-based rating scale used for assessment. Table~\ref{tab:instrument_nophase} reports the indicator-specific rating questions used by the judge.

\section{Activation and mediation pathways}

In the context of flood disasters, flood exposure is conceptualised as an environmental stressor that can activate interconnected psychological, social, and institutional resilience processes. Rather than treating flood-related wellbeing as a single outcome, we hypothesise that flood stressors activate multiple domains of wellbeing simultaneously and that affective--cognitive distress is further shaped through efficacy-related and social-protective pathways. Drawing on stress appraisal theory, self-efficacy theory, social-buffering theory, and disaster-resilience research, as shown in Figure~\ref{fig:wellbeing_hypotheses}, we propose the following hypotheses:

\begin{enumerate}[label={},leftmargin=3cm]
    
    \item[\textbf{\emph{Hypothesis I.}}]\label{HypothesisI}
    Flood stressors directly activate the three flood-wellbeing domains: BE$-$ (\emph{affective--cognitive distress appraisal}), DO$-$ (\emph{diminished perceived efficacy and agency}), and CONNECT$+$ (\emph{perceived relational--institutional connectedness}). This hypothesis tests whether climate stressors are associated not only with distress, but also with efficacy-related disruption and the mobilisation of relational and institutional protective resources.

    \item[\textbf{\emph{Hypothesis II.}}]\label{HypothesisII}
    DO$-$ mediates the activation of BE$-$ under flood stressors through an appraisal--efficacy activation pathway, represented by: \textbf{\emph{flood stressors} $\rightarrow$ DO$-$ $\rightarrow$ BE$-$}. This hypothesis tests whether flood-related deficits in coping efficacy, daily functioning, and access to resources are associated with heightened affective--cognitive distress.

    \item[\textbf{\emph{Hypothesis III.}}]\label{HypothesisIII}
    CONNECT$+$ and DO$-$ mediate the activation of BE$-$ under flood stressors through a social-buffering activation pathway, represented by: \textbf{\emph{flood stressors} $\rightarrow$ CONNECT$+$ $\rightarrow$ DO$-$ $\rightarrow$ BE$-$}. This hypothesis tests whether relational and institutional protective resources co-emerge with efficacy-related disruption and, through this pathway, contribute to the observed appraisal of flood-related adversity.
    
\end{enumerate}

As shown in Figure~\ref{fig:wellbeing_hypotheses}, there also allows for a mediation pathway from \textbf{\emph{flood stressors}} to \textbf{BE$-$} through \textbf{CONNECT$+$}. However, the empirical association for this pathway appears weaker than those observed for the appraisal--efficacy pathway, \textbf{\emph{flood stressors}} $\rightarrow$ \textbf{DO$-$} $\rightarrow$ \textbf{BE$-$}, and the extended social-buffering pathway, \textbf{\emph{flood stressors}} $\rightarrow$ \textbf{CONNECT$+$} $\rightarrow$ \textbf{DO$-$} $\rightarrow$ \textbf{BE$-$}. This pattern suggests that \textbf{CONNECT$+$} is not primarily expressed as a direct mediator of affective--cognitive distress appraisal. Rather, \textbf{CONNECT$+$} appears to operate as a relational--institutional resource mechanism that consolidates perceived coping efficacy and personal agency. In this interpretation, social support, protective mobilisation, and institutional confidence mediate flood-related appraisal primarily through \textbf{DO$-$}, by reducing efficacy deficits and strengthening agency under flood exposure.

\begin{figure}
    \centering
    \subfloat[Post length distribution.]{%
        \includegraphics[width=0.48\linewidth]{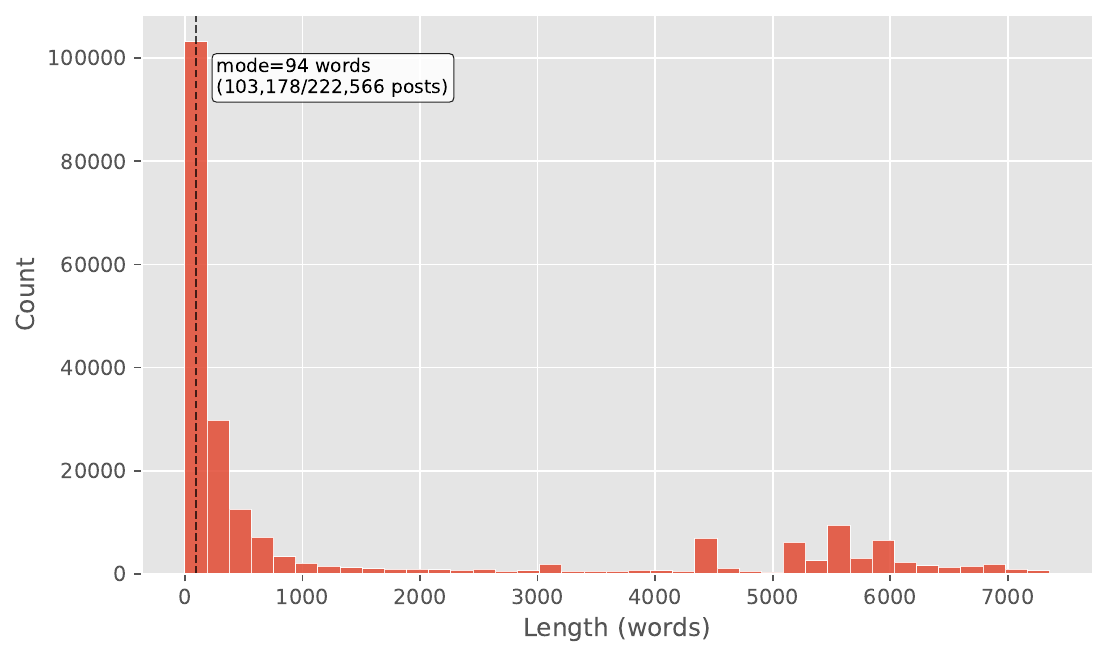}%
        \label{fig:post_length}%
    }
    \subfloat[Post count per year.]{%
        \includegraphics[width=0.48\linewidth]{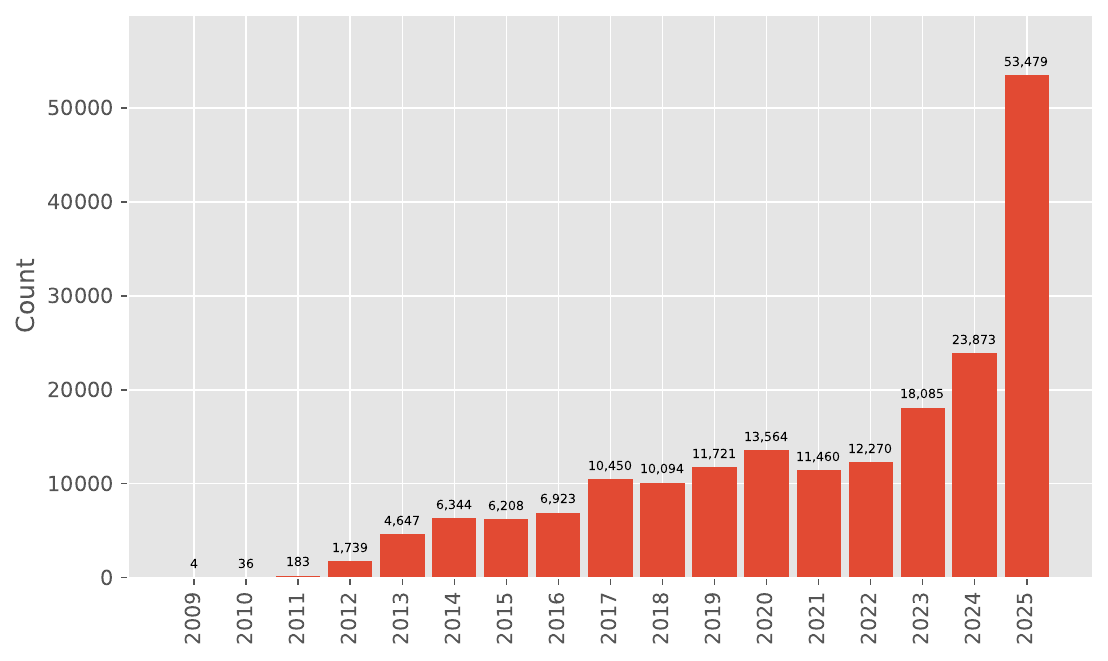}%
        \label{fig:posts_over_time}%
    }
    
    \caption{\textbf{Discourse Statistics.} The post-length distribution has mean $1{,}578$, median $223$, and mode $94$. The annual post count has mean $11{,}240$, median $10{,}094$, and maximum $53{,}479$, occurring on the year 2025.}
    
    \label{fig:mcl_stats}
    
\end{figure}

\section{Data}

\paragraph{\textbf{MCL--Irish--Flood--Discourse corpus}.} The Meta Content Library (MCL) is a research platform developed by Meta that provides approved researchers with access to publicly available content and associated metadata from Meta platforms for large-scale social and behavioural research \citep{Meta_MCL_v6}. Our research dataset, referred to as \emph{MCL--Irish--Flood--Discourse}, was collected through the MCL research platform. Fluvial and pluvial floods are typically embedded within broader hydro-meteorological processes characterised by elevated precipitation and storm activity. To capture discourse concerning both flood impacts and the meteorological conditions associated with flood occurrence, data retrieval employed the keywords \textit{flood}, \textit{rain}, and \textit{storm}, combined with at least one Irish geographic marker (\eg, Ireland, Cork, Galway, Dublin, \etc). The resulting corpus comprises approximately 24{,}000$+$ Facebook and Instagram posts related to flood events in Ireland. The dataset spans 2012--2025. For each record, Meta Content Library provides a unique post identifier, post content, creation timestamp, and associated platform-specific engagement and reaction metrics.

\begin{figure}
    \centering

    \resizebox{0.7\linewidth}{!}{
        \begingroup
\definecolor{cardgreen}{HTML}{6FA86B}
\definecolor{cardgreenfill}{HTML}{F6FAF4}
\definecolor{icongreenbg}{HTML}{DCEBD6}
\definecolor{greentext}{HTML}{2E6B33}
\definecolor{greenarr}{HTML}{4E8C4A}
\definecolor{bluearr}{HTML}{2F6FE0}
\definecolor{bluetext}{HTML}{1F3A8A}
\definecolor{robotblue}{HTML}{3F6FD6}
\definecolor{circbluefill}{HTML}{EEF3FC}
\definecolor{circbluering}{HTML}{C7D8F2}
\definecolor{heartpurple}{HTML}{7E5BD0}
\definecolor{circpurplefill}{HTML}{F2EEFB}
\definecolor{circpurplering}{HTML}{D8CCEF}
\def\iconbrainheart#1{%
\begin{scope}[shift={#1}, scale=0.5, draw=greentext, line width=0.9pt, line cap=round, line join=round]
  \draw[fill=none] (0,0.55)
    .. controls (-0.5,0.8) and (-0.95,0.45) .. (-0.8,0.05)
    .. controls (-1.05,-0.25) and (-0.65,-0.6) .. (-0.32,-0.48)
    .. controls (-0.18,-0.7) and (0.18,-0.7) .. (0.32,-0.48)
    .. controls (0.65,-0.6) and (1.05,-0.25) .. (0.8,0.05)
    .. controls (0.95,0.45) and (0.5,0.8) .. (0,0.55) -- cycle;
  \draw (-0.5,0.45) .. controls (-0.3,0.3) .. (-0.5,0.1);
  \draw (0.5,0.45) .. controls (0.3,0.3) .. (0.5,0.1);
  \draw[fill=greentext] (0,0.12) .. controls (0.08,0.3) and (0.32,0.24) .. (0.3,0.04)
    .. controls (0.28,-0.1) and (0.05,-0.2) .. (0,-0.28)
    .. controls (-0.05,-0.2) and (-0.28,-0.1) .. (-0.3,0.04)
    .. controls (-0.32,0.24) and (-0.08,0.3) .. (0,0.12) -- cycle;
\end{scope}}
\def\iconclipboard#1{%
\begin{scope}[shift={#1}, scale=0.5, draw=greentext, line width=0.9pt, line cap=round, line join=round]
  \draw[rounded corners=2pt] (-0.55,-0.75) rectangle (0.55,0.75);
  \draw[fill=greentext, rounded corners=1pt] (-0.2,0.62) rectangle (0.2,0.86);
  \foreach \y in {0.35,0.05,-0.25}{
    \draw (-0.35,\y) rectangle (-0.18,\y+0.14);
    \draw[line width=0.6pt] (-0.32,\y+0.07) -- (-0.26,\y+0.01) -- (-0.20,\y+0.12);
    \draw (-0.05,\y+0.07) -- (0.38,\y+0.07);
  }
\end{scope}}
\def\iconrobot#1{%
\begin{scope}[shift={#1}, scale=0.55, draw=robotblue, line width=1.0pt, line cap=round, line join=round]
  \draw (0,0.95) -- (0,0.7); \fill[robotblue] (0,1.0) circle (0.07);
  \draw[fill=white, rounded corners=4pt] (-0.7,-0.55) rectangle (0.7,0.7);
  \draw[fill=white, rounded corners=1.5pt] (-0.86,-0.15) rectangle (-0.7,0.25);
  \draw[fill=white, rounded corners=1.5pt] (0.7,-0.15) rectangle (0.86,0.25);
  \fill[robotblue] (-0.3,0.18) circle (0.13);
  \fill[robotblue] (0.3,0.18) circle (0.13);
  \draw (-0.28,-0.22) .. controls (-0.1,-0.4) and (0.1,-0.4) .. (0.28,-0.22);
\end{scope}}
\def\iconheartpulse#1{%
\begin{scope}[shift={#1}, scale=0.68]
  \fill[heartpurple] (0,0.3) .. controls (0.35,0.75) and (1.0,0.35) .. (0.55,-0.2)
    .. controls (0.35,-0.45) and (0.05,-0.6) .. (0,-0.72)
    .. controls (-0.05,-0.6) and (-0.35,-0.45) .. (-0.55,-0.2)
    .. controls (-1.0,0.35) and (-0.35,0.75) .. (0,0.3) -- cycle;
  \draw[white, line width=1.0pt, line cap=round, line join=round]
    (-0.5,-0.05) -- (-0.22,-0.05) -- (-0.08,0.28) -- (0.1,-0.32) -- (0.24,-0.05) -- (0.5,-0.05);
\end{scope}}
\begin{tikzpicture}[>=Stealth, font=\sffamily,
  card/.style={rounded corners=14pt, draw=cardgreen, line width=1pt, fill=cardgreenfill,
               minimum width=4.6cm, minimum height=4.2cm},
  bcirc/.style={circle, line width=1.2pt, minimum size=2.6cm}]
\node[card] (C1) at (0,6) {};
\node[card] (C2) at (8,6) {};
\fill[icongreenbg] ($(C1.north)+(0,-1.0)$) circle (0.7);
\fill[icongreenbg] ($(C2.north)+(0,-1.0)$) circle (0.7);
\iconbrainheart{($(C1.north)+(0,-1.0)$)}
\iconclipboard{($(C2.north)+(0,-1.0)$)}
\node[greentext, font=\sffamily\bfseries\large, align=center] at ($(C1.center)+(0,-0.5)$) {Wellbeing\\ appraisal model};
\node[greentext, font=\sffamily\bfseries\large, align=center] at ($(C2.center)+(0,-0.5)$) {Wellbeing\\ instrument};
\foreach \c in {C1,C2}{
  \draw[cardgreen!60, line width=0.8pt] ($(\c.south)+(-0.7,0.6)$) -- ($(\c.south)+(-0.15,0.6)$);
  \fill[cardgreen] ($(\c.south)+(0,0.6)$) circle (0.07);
  \draw[cardgreen!60, line width=0.8pt] ($(\c.south)+(0.15,0.6)$) -- ($(\c.south)+(0.7,0.6)$);
}
\node[bcirc, draw=circbluering, fill=circbluefill] (FB) at (0,0) {};
\node[bcirc, draw=circbluering, fill=circbluefill] (LLM) at (8,0) {};
\node[bcirc, draw=circpurplering, fill=circpurplefill] (WS) at (16,0) {};
\draw[circbluering, line width=0.8pt, dotted] (FB) circle (1.55);
\draw[circbluering, line width=0.8pt, dotted] (LLM) circle (1.55);
\draw[circpurplering, line width=0.8pt, dotted] (WS) circle (1.55);
\iconrobot{(LLM.center)}
\iconheartpulse{(WS.center)}
\begin{scope}[shift={(FB.center)}, scale=0.55, draw=robotblue, line width=1.0pt, line join=round]
  \fill[robotblue, rounded corners=5pt] (-0.7,-0.5) rectangle (0.7,0.55);
  \fill[robotblue] (-0.45,-0.45) -- (-0.15,-0.45) -- (-0.45,-0.85) -- cycle;
  \draw[white, line width=1.4pt, line cap=round] (-0.35,0.15) -- (0.35,0.15);
  \draw[white, line width=1.4pt, line cap=round] (-0.35,-0.12) -- (0.15,-0.12);
\end{scope}
\node[bluetext, font=\sffamily\bfseries\large, align=center] at ($(FB.south)+(0,-0.85)$) {Curated public\\ discourse};
\node[bluetext, font=\sffamily\bfseries\large, align=center] at ($(LLM.south)+(0,-0.85)$) {LLM};
\node[bluetext, font=\sffamily\bfseries\large, align=center] at ($(WS.south)+(0,-0.85)$) {Wellbeing\\ scores};
\draw[->, line width=1.6pt, draw=greenarr] ($(C1.east)+(0,0.6)$) -- ($(C2.west)+(0,0.6)$);
\draw[->, line width=1.6pt, draw=greenarr] (C2.south) -- (LLM.north);
\draw[->, line width=1.4pt, draw=bluearr] (FB.east) -- (LLM.west);
\draw[->, line width=1.4pt, draw=bluearr] (LLM.east) -- (WS.west);
\foreach \x in {0,1}{
  \begin{scope}[shift={($(FB.east)+(0.25+0.18*\x,0)$)}]
    \draw[bluearr, line width=1.2pt, line cap=round] (-0.06,0.14) -- (0.08,0) -- (-0.06,-0.14);
  \end{scope}
  \begin{scope}[shift={($(LLM.east)+(0.25+0.18*\x,0)$)}]
    \draw[bluearr, line width=1.2pt, line cap=round] (-0.06,0.14) -- (0.08,0) -- (-0.06,-0.14);
  \end{scope}
}
\end{tikzpicture}
\endgroup
    }
    
    \caption{\textbf{Wellbeing Reasoning Pipeline.} The wellbeing reasoning pipeline consists of three components: a bespoke flood-wellbeing instrument, a curated public-discourse corpus, and an LLM reasoning engine. Grounded in theories of wellbeing appraisal, coping efficacy, and social buffering, the instrument defines the wellbeing domains, indicators, and evidence-based rating scales that serve as the rubrics for constrained discourse assessment. Public discourse collected from MCL is curated through preprocessing and outlier filtering, including the removal of the upper 5\% quantile of post length. The curated discourse records are subsequently evaluated by the LLM through constrained wellbeing reasoning against the instrument definitions to produce multidimensional flood-wellbeing scores at scale.}
    
    \label{fig:wellbeing_reasoning_pipeline}
\end{figure}
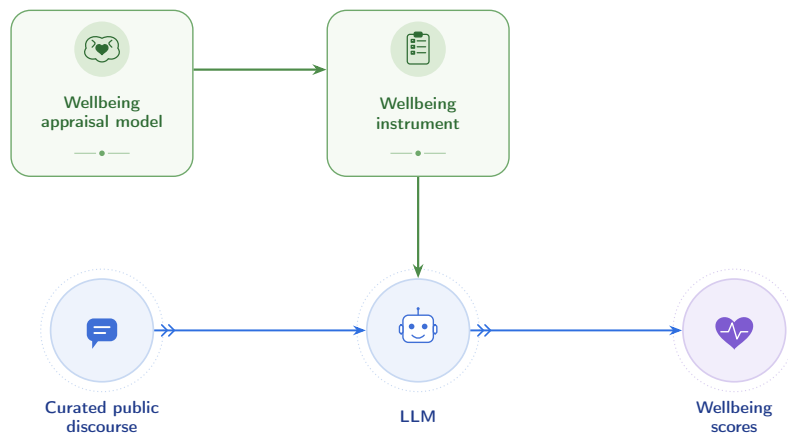

\begin{figure}
    \centering
    \includegraphics[width=0.9\linewidth]{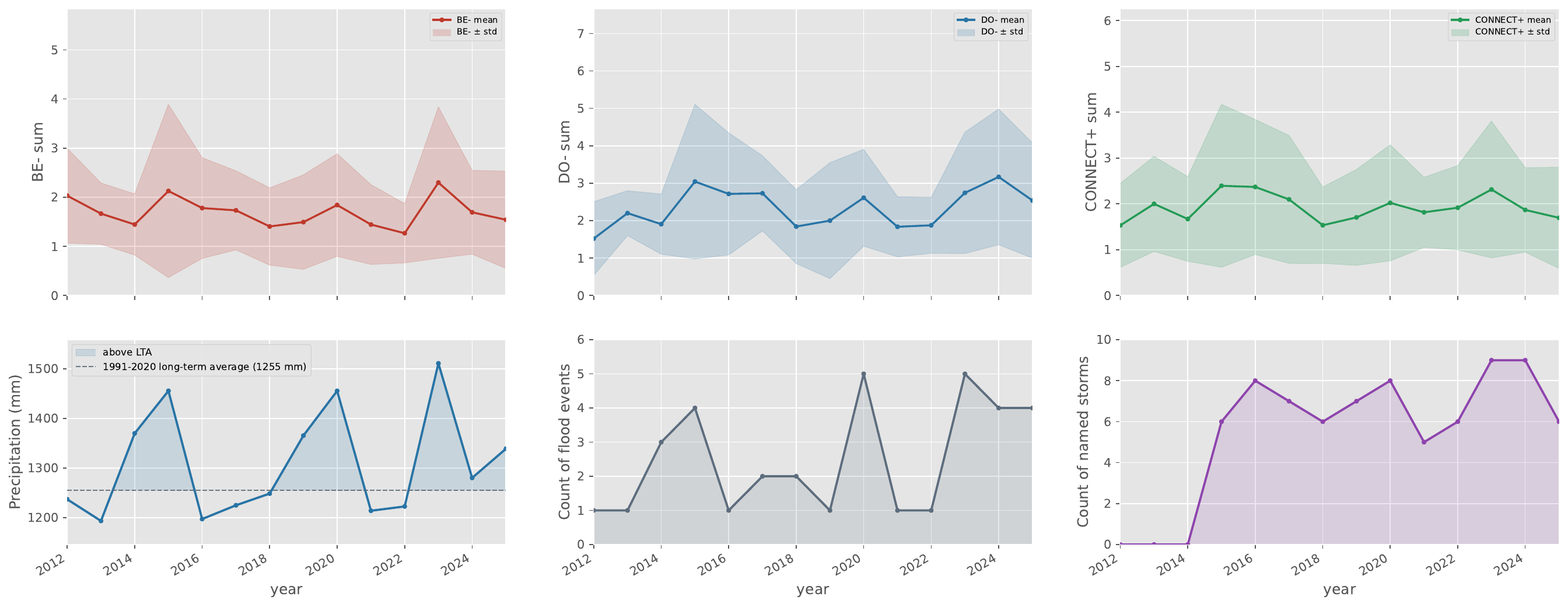}
    
    \caption{\textbf{Climate Stressors Activate Flood-Related Appraisal Processes.} The first row shows aggregated domain-level wellbeing scores, obtained by summing indicator scores within each domain. The second row presents the climate reference series, including precipitation volume, flood-event counts, and named-storm counts in Ireland from 2012 to 2025. Strong correlations are observed between the wellbeing domains and the climate reference series, while the wellbeing domains also exhibit substantial inter-domain correlations.}
    
    \label{fig:wellbeing_scores_and_climate_reference}
\end{figure}

\paragraph{\textbf{Climate reference data.}} To examine the relationship between climate hazards and flood-related wellbeing, we compile three climate reference datasets. These reference series are shown alongside the wellbeing indicators in Figure~\ref{fig:wellbeing_scores_and_climate_reference}. Details of each climate reference dataset are described below:
\begin{enumerate}

    \item \textbf{Precipitation.} Monthly precipitation records were obtained from the Central Statistics Office (CSO) \texttt{PxStat MTM01} monthly rainfall series for the Met \'Eireann synoptic station network \citep{CSO_MTM01}. The mean of monthly rainfall across the 15-station network was multiplied by a calibration factor of $1.139$ to produce a gridded-equivalent national precipitation series aligned with the gridded annual totals published by Met \'Eireann. The factor is the ratio of the Met \'Eireann 1991--2020 gridded annual long-term average (LTA), $1{,}255$~mm, to the corresponding 15-station network mean of station-level 1991--2020 annual means ($1{,}101.6$~mm): $1{,}255 / 1{,}101.6 \approx 1.139$. The network mean is systematically lower because the synoptic stations cluster in lowland, coastal, and population-centre locations and thereby under-sample the wet Atlantic uplands and western mountains that are captured by the spatially-interpolated gridded product; the scalar correction restores the national-scale total without introducing a second model. For years not covered by CSO MTM01 at the time of analysis (from 2023 onwards), annual gridded precipitation totals from Met \'Eireann \citep{MetEireann2024_AnnualClimate} were apportioned across months using the 1991--2020 climatological seasonal cycle; the same template was used to fill any missing monthly values within CSO-covered years. The 1991--2020 LTA annual gridded precipitation reference adopted in this study is $1{,}255$~mm (the full annual precipitation series is provided in Table~\ref{tab:appendix_precip} of Appendix~\ref{app:precip}).

    \item \textbf{Named storm.} The 78 named storms used as event reference points are compiled from the Met \'Eireann Storm Centre seasonal pages \citep{MetEireann_StormCentre} and cover the 2015--2016 through 2025--2026 storm seasons. Each entry is encoded with its calendar year and the month in which peak winds occurred over Ireland. The joint UK / Ireland storm-naming scheme launched in November 2015, so the series is structurally zero before that date; the Royal Netherlands Meteorological Institute (KNMI) joined the partnership from the 2019--2020 season. The list also includes ex-tropical and ex-hurricane systems re-named by Met \'Eireann when they affected Ireland (e.g.\ Ophelia 2017, Lorenzo 2019) and Iberian-named storms from the Southwestern Group (Portugal / Spain / France / Belgium / Luxembourg) with recorded Irish landfall impact (\eg, Jorge 2020, Antoni 2023, Claudia 2025); storms with no Irish landfall are omitted (see Table~\ref{tab:appendix_storms} of Appendix~\ref{app:storms}).

    \item \textbf{Dated flood event.} The 31 major flood events used as event reference points were curated from \citet{OPW_Floodinfo}, RT\'E and Irish Examiner news archives, FloodList, and Met \'Eireann post-event summaries, and span 2011--2026 (so the flood record begins four years earlier than the named-storm record). Each entry is encoded with its calendar year, month, and a multiplicity count that allows a single calendar month to carry several distinct sub-events (\eg, December 2015, count$=$4, bundles Storms Desmond, Eva, and Frank with the protracted Shannon-basin flooding they generated). The multiplicity counts are best-effort and illustrative; the qualitative month-of-occurrence pattern is the primary signal carried into the downstream analysis (see Table~\ref{tab:appendix_floods} of Appendix~\ref{app:floods}).
    
\end{enumerate}

\begin{figure}
    \centering
    
  \begin{minipage}{0.4\textwidth}
    \centering

    \resizebox{0.8\linewidth}{!}{
        \begin{tabular}{@{\hspace{2\tabcolsep}}lrr@{\hspace{2\tabcolsep}}}
        \toprule
        \textbf{Keyword} & \textbf{Count of posts} & \textbf{Count of users} \\
        \midrule
        \textit{flood} & 9{,}442 & 1{,}092 \\
        \textit{rain}  & 8{,}731 & 1{,}364 \\
        \textit{storm} & 6{,}479 & 632 \\
        \midrule
         \textit{total}  & 24{,}652 & 2{,}297\\
        \bottomrule
        \end{tabular}\label{fig:scoring_stats:keywords}
     }

     \medskip
     {\footnotesize\textbf{Note.} Total post counts are reported before top-5\% post-length filtering.}

  \end{minipage}
  \begin{minipage}{0.55\textwidth}
    \centering
    \includegraphics[width=1\linewidth]{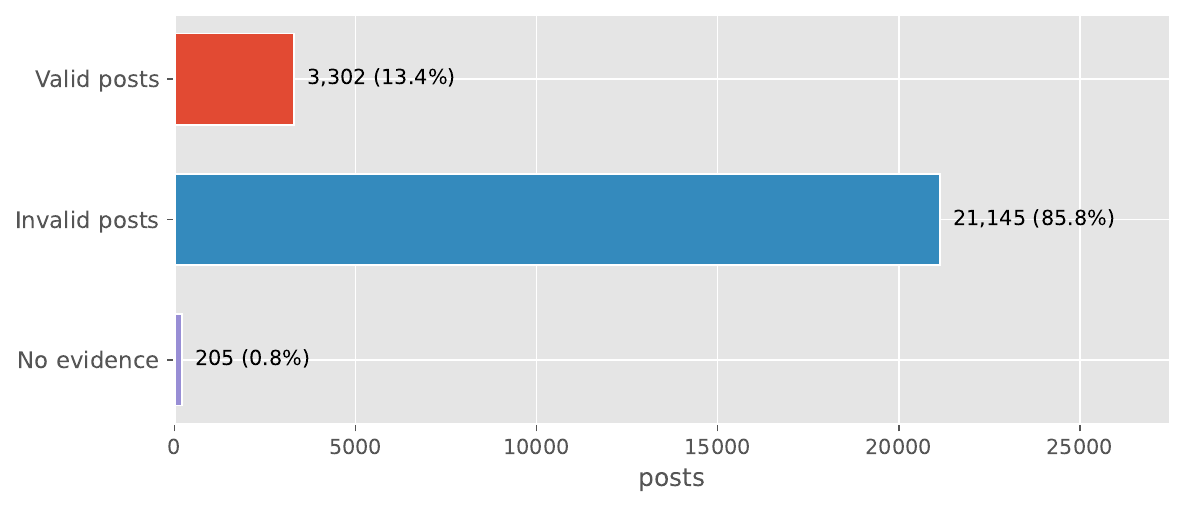}
    \label{fig:scoring_stats:hist}
  \end{minipage}


    \caption{\textbf{Scoring Statistics.} Posts are analysed using the LLM (Qwen2.5-14B-Instruct). Of all processed posts, 13.4\% passed the flood-wellbeing relevance and evidence gate and were retained for scoring, 85.8\% were excluded as non-relevant records, and 0.8\% contained insufficient evidence for wellbeing assessment.}
    
    \label{fig:scoring_stats}

\end{figure}

\section{Wellbeing reasoning}

\paragraph{\textbf{Large language models for reasoning.}} Large language models (LLMs) have emerged as a powerful class of foundation models for natural-language understanding and generation. Early work demonstrated that scaling transformer-based language models to very large parameter counts yields strong few-shot and in-context learning capabilities across a wide range of tasks \citep{Brown2020_GPT3}. Subsequent advances, including instruction tuning and reinforcement learning from human feedback, further improved alignment with user intent and task-oriented reasoning performance \citep{Ouyang2022_InstructGPT, OpenAI2023_GPT4}. Open-weight model families, including the Qwen series used in this study, have extended these capabilities while enabling reproducible research deployment \citep{Qwen25_2024}. Beyond text generation, LLMs have demonstrated the ability to perform complex reasoning when guided by structured prompting strategies. Chain-of-thought prompting encourages models to articulate intermediate reasoning steps before producing a judgement, substantially improving performance on reasoning-intensive tasks \citep{Wei2022_ChainOfThought}, while related approaches such as zero-shot chain-of-thought prompting and self-consistency decoding further enhance reasoning reliability \citep{Kojima2022_ZeroShotCoT, Wang2023_SelfConsistency}. These developments motivate the use of LLMs as constraint-guided raters for extracting flood-wellbeing indicators from large-scale unstructured public discourse.

\paragraph{\textbf{Wellbeing reasoning pipeline.}}
As shown in Figure~\ref{fig:wellbeing_reasoning_pipeline}, we employ a large language model (LLM) as a reasoning engine for large-scale wellbeing assessment of flood-related public discourse. The reasoning pipeline comprises three components: a bespoke flood-wellbeing instrument, a curated discourse corpus, and an LLM-based reasoning framework. The instrument defines the wellbeing domains, indicators, and evidence-based rating scales, grounded in theories of affective appraisal, coping efficacy, and social buffering, which together serve as the assessment rubric for constrained wellbeing reasoning. Public discourse records collected from Meta Content Library (MCL) are first curated through preprocessing and outlier filtering, including the removal of posts within the upper 5\% quantile of post length. The curated records are subsequently evaluated by the LLM against the instrument definitions and scoring criteria, enabling structured and theory-informed extraction of multidimensional flood-wellbeing indicators from large-scale unstructured discourse. 

\begin{figure}
    \centering
    
    \includegraphics[width=0.9\linewidth]{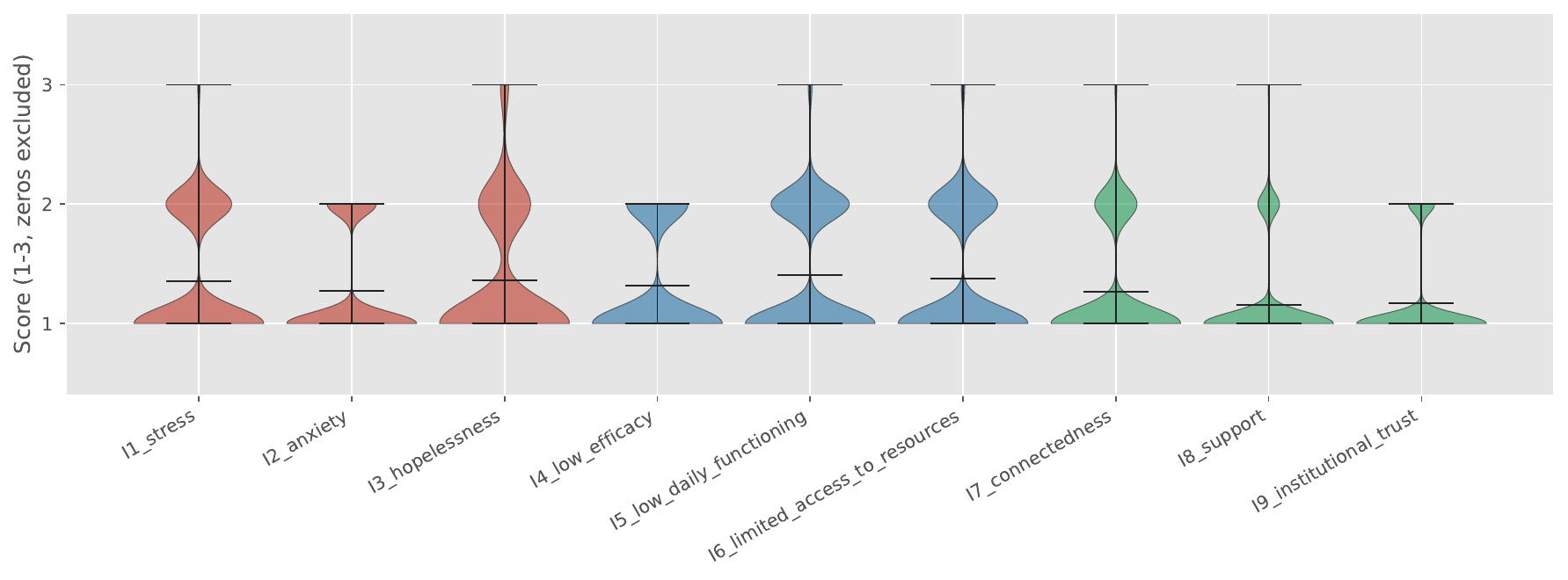}

    \caption{\textbf{Density Distribution of Scores.} The violin plots indicate that score~1 constitutes the dominant category across most indicators. The indicators \emph{I2: Anxiety}, \emph{I4: Low efficacy}, and \emph{I9: Institutional trust} exhibit no observations assigned the maximum score of~3.}
    
    \label{fig:indicator_dist_violins}
\end{figure}

\paragraph{\textbf{Aggregating domain scores.}}
Let $r$ index discourse records and $i$ index wellbeing indicators. Each flood-wellbeing domain $\mathcal{D}\in\{\mathrm{BE}{-},\,\mathrm{DO}{-},\,\mathrm{CONNECT}{+}\}$ consists of three indicators; for example, $\mathrm{BE}{-}=\{\mathrm{I1},\mathrm{I2},\mathrm{I3}\}$. For each record $r$ and each indicator $i\in\mathcal{D}$, the judge assigns an ordinal evidence score $x_{r,i}\in\{0,1,2,3\}$, where $x_{r,i}=0$ denotes the absence of admissible evidence for indicator $i$ in record $r$. We write $\tau(r)$ for the calendar year in which record $r$ was created. The domain score of record $r$, denoted $S^{\mathcal{D}}_{r}$, is defined as the sum of the three indicator scores belonging to that domain,
\begin{equation}
S^{\mathcal{D}}_{r}=\sum_{i\in\mathcal{D}}x_{r,i}\in\{0,1,\dots,9\},
\end{equation}
where an indicator with $x_{r,i}=0$ contributes zero to the sum. For a given year $t$, let $\mathcal{R}^{\mathcal{D}}_{t}$ denote the set of records created in that year whose domain score is strictly positive, that is, all records $r$ such that $\tau(r)=t$ and $S^{\mathcal{D}}_{r}>0$. Let $n^{\mathcal{D}}_{t}$ denote the number of records in $\mathcal{R}^{\mathcal{D}}_{t}$. Records with $S^{\mathcal{D}}_{r}=0$ contain no admissible evidence for the domain and are therefore excluded from the domain-level aggregation. The aggregated domain score $\bar{S}^{\mathcal{D}}_{t}$ and its dispersion $\sigma^{\mathcal{D}}_{t}$ are respectively defined as the arithmetic mean and population standard deviation of $S^{\mathcal{D}}_{r}$ over $\mathcal{R}^{\mathcal{D}}_{t}$,
\begin{equation}
\bar{S}^{\mathcal{D}}_{t}=\frac{1}{n^{\mathcal{D}}_{t}}
\sum_{r\in\mathcal{R}^{\mathcal{D}}_{t}}S^{\mathcal{D}}_{r},
\qquad
\sigma^{\mathcal{D}}_{t}=\sqrt{\frac{1}{n^{\mathcal{D}}_{t}}
\sum_{r\in\mathcal{R}^{\mathcal{D}}_{t}}
\bigl(S^{\mathcal{D}}_{r}-\bar{S}^{\mathcal{D}}_{t}\bigr)^{2}}.
\end{equation}
Summing the indicator scores within each record, rather than averaging the per-indicator means, weights records according to the total amount of evidence expressed for the domain, so that records in which several facets of the domain are jointly activated contribute proportionally more to the aggregate. Restricting the average to records with $S^{\mathcal{D}}_{r}>0$ follows from the evidence-gated scoring design: $x_{r,i}=0$ marks the absence of admissible evidence rather than a measured low intensity. Retaining zero-domain records would therefore conflate non-expression with low expression and would bias the estimate toward how frequently the domain is mentioned rather than how intensely it is expressed when present. We report $\bar{S}^{\mathcal{D}}_{t}\pm\sigma^{\mathcal{D}}_{t}$ for each domain $\mathcal{D}$ and year $t$.

\paragraph{\textbf{Treatment of zero-evidence scores.}} Social-media posts are often brief and may not contain observable evidence for every wellbeing indicator. Under the evidence-based scoring scheme, a score of zero is assigned only when no admissible evidence---including direct statements, descriptive content, tonal or pragmatic cues, figurative language, framing, relaying, silence, or satire---can be identified for a given indicator. A zero score therefore reflects the absence of evidence in the discourse record rather than the absence of the underlying construct and is treated as missing data in subsequent analyses; the lowest substantive score is~1. Accordingly, zero-valued indicator scores are excluded from indicator-level activation analyses, and posts whose domain score is zero (\ie, no indicator within the domain exhibits evidence) are excluded from the computation of domain-level temporal summaries.

\section{Results}
\label{sec:results}

\paragraph{\textbf{Scoring statistics.}} Figure~\ref{fig:scoring_stats} summarises the outcome of the wellbeing reasoning pipeline. Of the approximately 222{,}566 discourse records processed by the LLM, 13.4\% passed the flood-wellbeing relevance and evidence gate and were retained for scoring, whereas 85.8\% were excluded as non-relevant records and a further 0.8\% contained insufficient evidence for wellbeing assessment. The relatively low retention rate reflects the stringent relevance and evidence criteria imposed by the bespoke instrument, which requires observable evidence of flood-related wellbeing constructs before a score can be assigned. Figure~\ref{fig:indicator_dist_violins} presents the distribution of indicator scores among retained records. Across all indicators, score~1 is the dominant category, indicating that low-intensity evidence is the most common form of flood-related wellbeing expression in public discourse. Higher-intensity scores occur less frequently across all indicators, with substantial variation in their distributions. Notably, the indicators \emph{I2: Anxiety}, \emph{I4: Low efficacy}, and \emph{I9: Institutional trust} contain no observations assigned the maximum score of~3, suggesting that highly explicit expressions of severe anxiety, pronounced efficacy loss, and exceptionally strong institutional trust are comparatively uncommon within the discourse corpus. The full per-year wellbeing scores are provided in Table~\ref{tab:appendix_wellbeing}--\ref{tab:appendix_wellbeing_domains} of Appendix~\ref{app:wellbeing}.

\paragraph{\textbf{Activation of flood--wellbeing domain.}} 
Figure~\ref{fig:wellbeing_scores_and_climate_reference} shows the temporal evolution of the three wellbeing domains alongside the climate reference series from 2012 to 2025. Across the observation period, the domain scores exhibit substantial co-movement with one another and broadly track periods of elevated precipitation, major flood events, and named-storm activity. Prominent increases are visible during flood-active years, including the 2015--2016 period associated with Storms Desmond, Eva, and Frank and the prolonged Shannon-basin flooding; the 2020 period associated with Storms Ciara, Dennis, and Ellen and the Skibbereen/Bandon flooding; the 2023 period associated with Storm Babet and the Midleton flooding; and the 2024--2025 period characterised by multiple severe storm events. In each of these exposure windows, BE$-$, DO$-$, and CONNECT$+$ rise concurrently rather than moving in opposing directions. The temporal co-activation of distress, efficacy-related deficits, and protective social resources is consistent with the joint activation of appraisal--efficacy and social-buffering pathways under recurrent flood exposure. Notably, higher levels of connectedness, support, and institutional trust do not coincide with lower levels of distress or efficacy-related impacts at the aggregate population level; instead, protective resources tend to increase alongside indicators of adversity. This pattern is more consistent with resource mobilisation and altruistic-community dynamics \citep{KaniastyNorris1995_AltruisticCommunity, Hobfoll1989_COR} than with the classical stress-buffering expectation that social resources suppress distress under conditions of elevated exposure \citep{CohenWills1985_SupportBuffering}.

\begin{figure}
    \centering
    \includegraphics[width=0.7\linewidth]{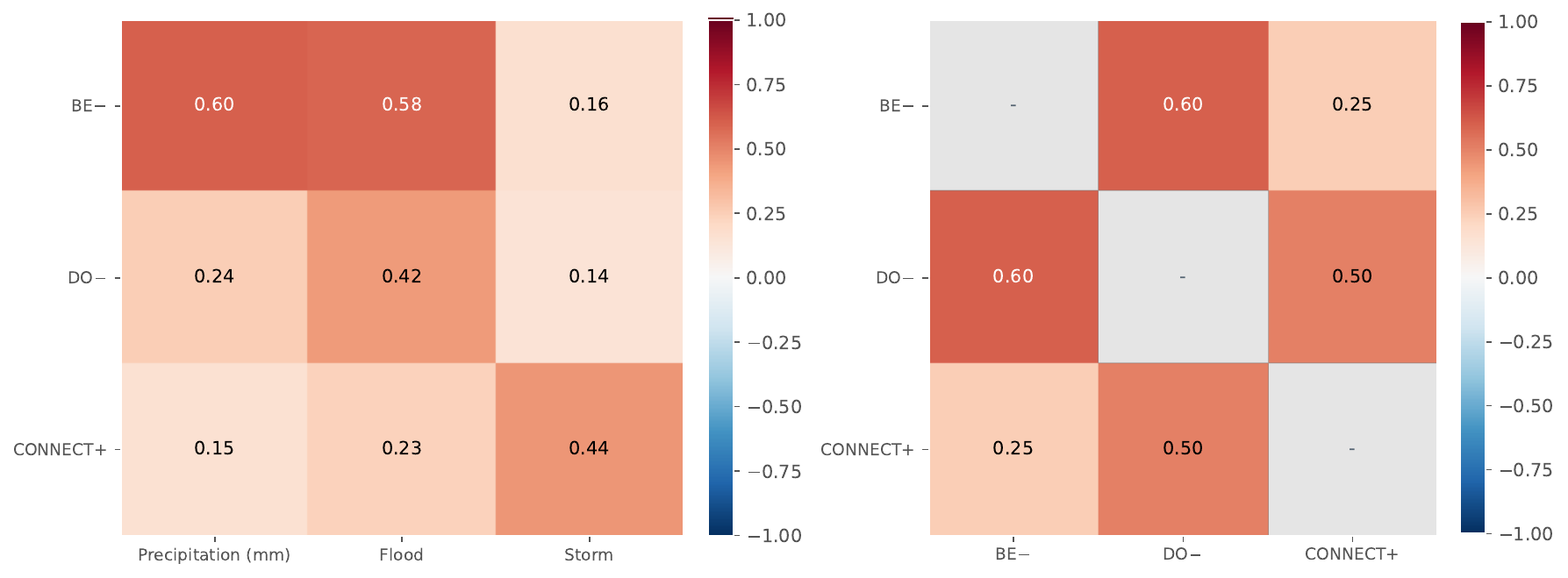}
    
    \caption{\textbf{Pearson Correlations of Domain--Climate and Domain--Domain Relationships.} Pearson correlation coefficients are computed between aggregated wellbeing domain scores and climate reference series, and between the wellbeing domains themselves. The resulting correlation structure provides evidence for the co-activation of appraisal--efficacy and social-buffering pathways under recurrent flood exposure.}
    
    \label{fig:pearson_correlations}
\end{figure}

\paragraph{\textbf{Mediation of flood--wellbeing domain.}} 
Figure~\ref{fig:pearson_correlations} quantifies the relationships between wellbeing domains and climate stressors, and between the wellbeing domains themselves. Among the climate reference series, precipitation exhibits the strongest association with wellbeing, showing moderate positive correlations with BE$-$ ($r=0.60$) and DO$-$ ($r=0.24$), and a weaker association with CONNECT$+$ ($r=0.15$). Flood-event frequency displays a similar pattern, correlating positively with BE$-$ ($r=0.58$), DO$-$ ($r=0.42$), and CONNECT$+$ ($r=0.23$). Named storms are most strongly associated with CONNECT$+$ ($r=0.44$), while their correlations with BE$-$ ($r=0.16$) and DO$-$ ($r=0.14$) are comparatively weak. The domain--domain correlations reveal substantial co-activation among the three wellbeing domains. BE$-$ and DO$-$ exhibit the strongest association ($r=0.60$), consistent with an appraisal--efficacy pathway in which heightened distress is accompanied by diminished efficacy and adaptive capacity. DO$-$ and CONNECT$+$ are also positively correlated ($r=0.50$), indicating that protective social and institutional resources tend to become more salient during periods of elevated efficacy-related adversity. The weaker but still positive correlation between BE$-$ and CONNECT$+$ ($r=0.25$) further suggests that social support and institutional trust are activated alongside, rather than in opposition to, flood-related distress. Taken together, these correlation patterns provide evidence for the co-emergence of appraisal--efficacy and social-buffering pathways under recurrent flood exposure.

\section{Discussion}

\paragraph{\textbf{Validation of Hypothesis I: direct activation pathway.}} Figure~\ref{fig:pearson_correlations} shows that climate stressors are positively associated with all three wellbeing domains, indicating that flood exposure directly activates affective appraisal, efficacy-related functioning, and social-protective processes. The strongest activation is observed for the BE$-$ domain, which exhibits moderate correlations with both precipitation ($r=0.60$) and flood-event frequency ($r=0.58$). This pattern is consistent with transactional theories of stress, which conceptualise environmental hazards as stressors that trigger threat appraisal, uncertainty, and emotional distress \citep{LazarusFolkman1984_StressAppraisal}. The DO$-$ domain also exhibits positive correlations with precipitation ($r=0.24$) and flood frequency ($r=0.42$), suggesting that periods of elevated flood activity are associated with greater expressions of efficacy loss, functional disruption, and resource constraints. This finding is consistent with resource-based theories of disaster adaptation, which emphasise the erosion of coping resources and adaptive capacity under prolonged environmental stress \citep{Hobfoll1989_COR, Norris2008_CommunityResilience}. The CONNECT$+$ domain displays comparatively weaker but consistently positive associations with precipitation ($r=0.15$) and flood frequency ($r=0.23$), while exhibiting its strongest association with named-storm activity ($r=0.44$). One possible explanation is that named storms are highly visible public events that attract intensive media attention, emergency communication, and community mobilisation, thereby increasing the salience of social support and institutional response within public discourse. Taken together, these results suggest that flood-related climate stressors do not activate distress alone; they simultaneously activate appraisal, efficacy, and social-protective processes, providing empirical evidence that flood exposure engages multiple dimensions of social--ecological resilience rather than a single pathway of psychological impact.

\paragraph{\textbf{Validation of Hypothesis II \& III: social--ecological resilience structure.}} Social--ecological resilience refers to the interconnected system of psychological, social, and institutional processes through which individuals and communities respond to, adapt to, and recover from environmental stressors \citep{Norris2008_CommunityResilience, Hobfoll1989_COR}. Rather than viewing disaster impacts as a single pathway from exposure to distress, the resilience perspective conceptualises adaptation as a structured network of interacting mechanisms linking affective appraisal, coping efficacy, social support, and institutional resources. Beyond the direct activation effects of climate stressors, the correlation structure in Figure~\ref{fig:pearson_correlations} suggests the presence of two interconnected resilience pathways. The first is an appraisal--efficacy pathway, represented by the strong positive association between DO$-$ and BE$-$ ($r=0.60$). Under this pathway, flood exposure activates efficacy-related deficits, including reduced coping capacity, impaired daily functioning, and constrained access to resources, which are in turn associated with heightened stress, anxiety, and hopelessness. This interpretation is consistent with transactional stress theory and disaster self-efficacy research, which argue that perceived coping capacity is a key determinant of emotional adjustment under disaster exposure \citep{LazarusFolkman1984_StressAppraisal, Bandura1977_SelfEfficacy, Benight1999b_HurricaneAndrewModel}. The second is a social-buffering pathway, represented by the positive association between CONNECT$+$ and DO$-$ ($r=0.50$), together with the indirect linkage between CONNECT$+$ and BE$-$ through DO$-$. This pattern suggests that periods characterised by greater efficacy-related adversity are also periods in which social connectedness, protective support, and institutional trust become more salient within public discourse. The pathway is consistent with theories of social support mobilisation, community resilience, and resource conservation, which propose that social and institutional resources constitute important protective capacities during disaster response and recovery \citep{CohenWills1985_SupportBuffering, KaniastyNorris1995_AltruisticCommunity, Hobfoll1989_COR, Norris2008_CommunityResilience}.

\paragraph{\textbf{Understanding the role of institutional support.}} Although a direct CONNECT$+$~$\rightarrow$~BE$-$ association is observable ($r=0.25$), it is substantially weaker than the DO$-$~$\rightarrow$~BE$-$ and CONNECT$+$~$\rightarrow$~DO$-$ relationships. This pattern suggests that perceived relational and institutional support is not primarily expressed in discourse as an immediate reduction of affective--cognitive distress. Rather, CONNECT$+$ appears to operate through the domain of efficacy and agency: institutional response, infrastructure protection, emergency services, community assistance, and access to support are more directly related to whether individuals perceive that they can cope, maintain functioning, and manage flood-related disruption. In this interpretation, institutional and social resources mediate flood-related appraisal mainly by reducing efficacy deficits and strengthening perceived agency, which in turn shapes the affective burden captured by BE$-$. Taken together, these findings support a social--ecological resilience structure in which climate stressors activate interconnected pathways of appraisal, efficacy, and social protection, consistent with mechanisms previously identified in disaster and resilience research.

\section{Conclusion}

This paper argues that flood-related wellbeing is not adequately understood as a single psychological outcome of disaster exposure. Drawing on flood-related public posts spanning 2012--2025, we show that flood stressors activate multiple interconnected wellbeing domains, including affective--cognitive distress, adaptive-capacity deficit, and relational--institutional recovery resources. These domains are further organised through appraisal--efficacy and social-buffering mediation pathways, revealing a coherent social--ecological resilience structure underlying how flood exposure is experienced and articulated in public discourse. Although the constituent mechanisms of appraisal, efficacy, social support, and resilience have been extensively studied in disaster psychology, they have typically been examined in isolation or through event-specific empirical studies. Our findings suggest that these mechanisms should instead be understood as components of an integrated resilience structure activated under recurrent flood exposure. More broadly, the study demonstrates that unobtrusive public discourse can provide a population-level and longitudinal perspective on wellbeing, adaptation, and resilience, complementing conventional survey-based approaches and enabling the investigation of psychosocial responses to climate-related hazards as they unfold over time.

\section*{Acknowledgments}


This publication was made possible from research conducted with the financial support of the Digi\'Eire research project at the University of Galway. The authors gratefully acknowledge the computational support provided by the Irish Centre for High-End Computing (ICHEC) and Luxembourg's national supercomputer (MeluXina). The authors also acknowledge the research data support provided by Meta Platforms, Inc. through Meta Content Library.


\bibliographystyle{cas-model2-names}

\bibliography{references}

\newpage
\appendix
\appendix

\definecolor{BeaufNG}{HTML}{DCE6F2}
\definecolor{BeaufG}{HTML}{BCD2EA}
\definecolor{BeaufSG}{HTML}{8FB4DC}
\definecolor{BeaufS}{HTML}{9B6FB8}
\definecolor{BeaufVS}{HTML}{6E2C91}
\definecolor{BeaufH}{HTML}{000000}

\section{Climate reference: annual national precipitation (2012--2025)}\label{app:precip}

\begin{center}

\captionof{table}{Annual national gridded-equivalent precipitation (mm).}

\label{tab:appendix_precip}

\resizebox{0.4\linewidth}{!}{
\begin{tabular}{@{\hspace{3\tabcolsep}}l l l@{\hspace{3\tabcolsep}}}
\toprule
\textbf{Year} & \textbf{Precipitation (mm)} & \textbf{Source} \\ \midrule
2012 & 1,236.9 & CSO \\
2013 & 1,193.3 & CSO \\
2014 & 1,369.8 & CSO \\
2015 & 1,455.5 & CSO \\
2016 & 1,197.1 & CSO \\
2017 & 1,224.8 & CSO \\
2018 & 1,248.4 & CSO \\
2019 & 1,365.3 & CSO \\
2020 & 1,455.4 & CSO \\
2021 & 1,213.8 & CSO \\
2022 & 1,222.5 & CSO \\
2023 & 1,510.9 & Met \'Eireann \\
2024 & 1,280.0 & Met \'Eireann \\
2025 & 1,338.8 & Met \'Eireann \\
\bottomrule
\end{tabular}
}

\medskip
{\footnotesize\textbf{Note.} Values for 2012--2022 are derived from the Central Statistics Office (CSO) \texttt{MTM01} 15-station network annual mean, scaled by the gridded calibration factor of $1.139$. Values for 2023--2025 are Met \'Eireann gridded annual totals. The 1991--2020 gridded long-term average is $1{,}255$~mm.}

\end{center}

\newpage
\section{Climate reference: named storm (2012--2025)}\label{app:storms}

{\footnotesize
\begin{longtable}{@{\hspace{2\tabcolsep}}l l l r@{\hspace{2\tabcolsep}}}

\caption{Named storms affecting Ireland during the 2015--2026 storm seasons. The joint Met Office--Met \'Eireann storm-naming scheme commenced in November 2015; consequently, no named storms are reported for 2012--2015. Maximum gust speeds and corresponding Beaufort-class colour codes (based on sustained 10-minute mean wind speeds) are obtained from the Met \'Eireann Storm Centre. Entries marked \emph{n/a} indicate storms for which no official Met \'Eireann extreme-wind summary is available. Beaufort sustained-wind classification: \colorbox{BeaufNG}{Near Gale}, \colorbox{BeaufG}{Gale}, \colorbox{BeaufSG}{Strong Gale}, \colorbox{BeaufS}{\textcolor{white}{Storm}}, \colorbox{BeaufVS}{\textcolor{white}{Violent Storm}}, \colorbox{BeaufH}{\textcolor{white}{Hurricane}}.}

\label{tab:appendix_storms}\\
\toprule
\textbf{Year} & \textbf{Date} & \textbf{Name} & \textbf{Highest gust (kn / km\,h$^{-1}$)} \\ \midrule
\endfirsthead
\toprule \textbf{Year} & \textbf{Date} & \textbf{Name} & \textbf{Highest gust (kn / km\,h$^{-1}$)} \\ \midrule \endhead
2015 & 12--13 Nov & \colorbox{BeaufS}{\textcolor{white}{Abigail}} & 65 / 120 \\
2015 & 17--18 Nov & \colorbox{BeaufS}{\textcolor{white}{Barney}} & 69 / 128 \\
2015 & 28--29 Nov & \colorbox{BeaufS}{\textcolor{white}{Clodagh}} & 65 / 120 \\
2015 & 4--6 Dec & \colorbox{BeaufSG}{Desmond} & 64 / 119 \\
2015 & 23--26 Dec & \colorbox{BeaufS}{\textcolor{white}{Eva}} & 73 / 135 \\
2015 & 29--30 Dec & \colorbox{BeaufS}{\textcolor{white}{Frank}} & 72 / 133 \\
2016 & 28--29 Jan & \colorbox{BeaufS}{\textcolor{white}{Gertrude}} & 70 / 130 \\
2016 & 1--2 Feb & \colorbox{BeaufS}{\textcolor{white}{Henry}} & 70 / 130 \\
2016 & 7--8 Feb & \colorbox{BeaufS}{\textcolor{white}{Imogen}} & 73 / 135 \\
2016 & 1--3 Mar & \colorbox{BeaufS}{\textcolor{white}{Jake}} & 72 / 133 \\
2016 & 27--28 Mar & \colorbox{BeaufG}{Katie} & 57 / 106 \\
2016 & 19--20 Nov & \colorbox{BeaufG}{Angus} & 44 / 81 \\
2016 & 23--26 Dec & \colorbox{BeaufG}{Barbara} & 60 / 111 \\
2016 & 25--26 Dec & \colorbox{BeaufG}{Conor} & 63 / 117 \\
2017 & 22--23 Feb & \colorbox{BeaufVS}{\textcolor{white}{Doris}} & 76 / 141 \\
2017 & 25--26 Feb & \colorbox{BeaufG}{Ewan} & 57 / 106 \\
2017 & 12--13 Sep & \colorbox{BeaufG}{Aileen} & 47 / 87 \\
2017 & 16 Oct & \colorbox{BeaufVS}{\textcolor{white}{Ophelia}} & 84 / 156 \\
2017 & 21 Oct & \colorbox{BeaufG}{Brian} & 62 / 115 \\
2017 & 6--7 Dec & \colorbox{BeaufG}{Caroline} & 54 / 100 \\
2017 & 30--31 Dec & \colorbox{BeaufS}{\textcolor{white}{Dylan}} & 67 / 124 \\
2018 & 2--3 Jan & \colorbox{BeaufVS}{\textcolor{white}{Eleanor}} & 84 / 156 \\
2018 & 16--17 Jan & \colorbox{BeaufS}{\textcolor{white}{Fionn}} & 74 / 137 \\
2018 & 1--3 Mar & \colorbox{BeaufG}{Emma} & 56 / 104 \\
2018 & 19 Sep & \colorbox{BeaufVS}{\textcolor{white}{Ali}} & 79 / 146 \\
2018 & 20--21 Sep & \colorbox{BeaufG}{Bronagh} & 44 / 81 \\
2018 & 12--13 Oct & \colorbox{BeaufS}{\textcolor{white}{Callum}} & 67 / 124 \\
2019 & 15--16 Jan & \colorbox{BeaufG}{Deirdre} & 58 / 107 \\
2019 & 8--9 Feb & \colorbox{BeaufS}{\textcolor{white}{Erik}} & 65 / 120 \\
2019 & 3 Mar & \colorbox{BeaufG}{Freya} & 59 / 109 \\
2019 & 12--13 Mar & \colorbox{BeaufS}{\textcolor{white}{Gareth}} & 70 / 130 \\
2019 & 26--27 Apr & \colorbox{BeaufS}{\textcolor{white}{Hannah}} & 66 / 122 \\
2019 & 3 Oct & \colorbox{BeaufS}{\textcolor{white}{Lorenzo}} & 58 / 107 \\
2019 & 8--9 Dec & \colorbox{BeaufS}{\textcolor{white}{Atiyah}} & 69 / 128 \\
2020 & 13--14 Jan & \colorbox{BeaufS}{\textcolor{white}{Brendan}} & 73 / 135 \\
2020 & 8--9 Feb & \colorbox{BeaufS}{\textcolor{white}{Ciara}} & 68 / 126 \\
2020 & 15--16 Feb & \colorbox{BeaufS}{\textcolor{white}{Dennis}} & 66 / 122 \\
2020 & 28--29 Feb & \colorbox{BeaufS}{\textcolor{white}{Jorge}} & 72 / 133 \\
2020 & 19--20 Aug & \colorbox{BeaufVS}{\textcolor{white}{Ellen}} & 77 / 143 \\
2020 & 25 Aug & \colorbox{BeaufG}{Francis} & 52 / 96 \\
2020 & 31 Oct & \colorbox{BeaufS}{\textcolor{white}{Aiden}} & 70 / 130 \\
2020 & 26--27 Dec & \colorbox{BeaufSG}{Bella} & 66 / 122 \\
2021 & 19--20 Jan & \colorbox{BeaufNG}{Christoph} & 42 / 78 \\
2021 & 7--9 Feb & \colorbox{BeaufG}{Darcy} & 45 / 83 \\
2021 & 30 Jul & Evert & 31 / 57 \\
2021 & 26--27 Nov & \colorbox{BeaufS}{\textcolor{white}{Arwen}} & 70 / 130 \\
2021 & 7--8 Dec & \colorbox{BeaufVS}{\textcolor{white}{Barra}} & 73 / 135 \\
2022 & 28--30 Jan & \colorbox{BeaufG}{Malik} & 58 / 107 \\
2022 & 30--31 Jan & \colorbox{BeaufG}{Corrie} & 53 / 98 \\
2022 & 16--17 Feb & \colorbox{BeaufG}{Dudley} & 68 / 126 \\
2022 & 18 Feb & \colorbox{BeaufVS}{\textcolor{white}{Eunice}} & 74 / 137 \\
2022 & 19--21 Feb & \colorbox{BeaufS}{\textcolor{white}{Franklin}} & 75 / 139 \\
2022 & 11--12 Nov & Claudio & n/a \\
2023 & 16--17 Feb & \colorbox{BeaufG}{Otto} & 52 / 96 \\
2023 & 5--6 Aug & \colorbox{BeaufG}{Antoni} & 55 / 102 \\
2023 & 27--28 Sep & \colorbox{BeaufSG}{Agnes} & 63 / 117 \\
2023 & 18--21 Oct & \colorbox{BeaufG}{Babet} & 50 / 93 \\
2023 & 1--2 Nov & \colorbox{BeaufNG}{Ciaran} & 43 / 80 \\
2023 & 13 Nov & \colorbox{BeaufSG}{Debi} & 62 / 115 \\
2023 & 9 Dec & \colorbox{BeaufSG}{Elin} & 55 / 102 \\
2023 & 10 Dec & \colorbox{BeaufS}{\textcolor{white}{Fergus}} & 63 / 117 \\
2023 & 27--28 Dec & \colorbox{BeaufSG}{Gerrit} & 64 / 119 \\
2024 & 2 Jan & Henk & 36 / 67 \\
2024 & 21--22 Jan & \colorbox{BeaufVS}{\textcolor{white}{Isha}} & 74 / 137 \\
2024 & 23 Jan & \colorbox{BeaufS}{\textcolor{white}{Jocelyn}} & 66 / 122 \\
2024 & 6 Apr & \colorbox{BeaufSG}{Kathleen} & 61 / 113 \\
2024 & 22 Aug & \colorbox{BeaufG}{Lilian} & 52 / 96 \\
2024 & 19--20 Oct & \colorbox{BeaufS}{\textcolor{white}{Ashley}} & 74 / 137 \\
2024 & 23--24 Nov & \colorbox{BeaufS}{\textcolor{white}{Bert}} & 62 / 115 \\
2024 & 27 Nov & Conall & 25 / 46 \\
2024 & 6--7 Dec & \colorbox{BeaufVS}{\textcolor{white}{Darragh}} & 76 / 141 \\
2025 & 24 Jan & \colorbox{BeaufH}{\textcolor{white}{\'Eowyn}} & 99.5 / 184 \\
2025 & 26--27 Jan & Herminia & n/a \\
2025 & 16 Feb & Ivo & n/a \\
2025 & 3 Oct & \colorbox{BeaufVS}{\textcolor{white}{Amy}} & 80 / 148 \\
2025 & 14 Nov & Claudia & n/a \\
2025 & mid Dec & Bram & n/a \\
\bottomrule
\end{longtable}
}

\newpage
\section{Climate reference: Irish flood event (2012--2025)}\label{app:floods}
{\footnotesize
\begin{longtable}{@{\hspace{4\tabcolsep}}l l p{2.7cm} c p{5.4cm}@{\hspace{4\tabcolsep}}}
\caption{Major Irish flood events, 2012--2025, curated from OPW Flood Information Portal, RT\'E and Irish Examiner archives, FloodList, and Met \'Eireann post-event summaries. Category is assigned from event characteristics: \emph{fluvial} = river overflow, \emph{coastal} = tidal/storm-surge, \emph{pluvial} = intense-rainfall surface or flash flooding. \emph{Count} is the event multiplicity (distinct sub-events bundled within one calendar month), matching the flood-event reference series in Figure~\ref{fig:wellbeing_scores_and_climate_reference}.}\\
\label{tab:appendix_floods}\\
\toprule
\textbf{Date} & \textbf{Flood category} & \textbf{Location} & \textbf{Count} & \textbf{Note} \\ \midrule
\endfirsthead
\toprule \textbf{Date} & \textbf{Flood category} & \textbf{Location} & \textbf{Count} & \textbf{Note} \\ \midrule \endhead
Jun 2012 & Fluvial & Cork city & 1 & River Lee burst its banks \\
Jan 2013 & Coastal & Dublin & 1 & Coastal storm-surge flooding \\
Jan 2014 & Coastal & West coast & 1 & Atlantic-storm coastal flooding (6--8 Jan) \\
Feb 2014 & Pluvial & Limerick / Cork / Kerry & 2 & Storm Darwin: rainfall and wind-driven flooding \\
Dec 2015 & Fluvial & Shannon basin & 4 & Storms Desmond, Eva \& Frank; prolonged Shannon flooding \\
Dec 2016 & Coastal & Cork city & 1 & Tidal / post-Storm Conor flooding (27--29 Dec) \\
Oct 2017 & Coastal & South \& west coast & 1 & Storm Ophelia coastal flooding \\
Nov 2017 & Pluvial & Inishowen, Donegal & 1 & Flash flooding from intense rainfall \\
Jan 2018 & Coastal & West \& south coast & 1 & Storm Eleanor coastal flooding \\
Mar 2018 & Pluvial & Nationwide & 1 & Storm Emma / ``Beast from the East thaw flooding \\
Dec 2019 & Fluvial & South / southwest & 1 & Storm Atiyah rainfall flooding \\
Feb 2020 & Fluvial & Shannon basin / Cork & 2 & Storms Ciara \& Dennis river flooding \\
Aug 2020 & Fluvial & Bandon / Skibbereen & 2 & Storms Ellen \& Francis river flooding \\
Oct 2020 & Pluvial & South / east & 1 & Storm Aiden rainfall flooding \\
Dec 2021 & Pluvial & South / west & 1 & Storm Barra rainfall flooding \\
Feb 2022 & Coastal & North \& west coast & 1 & Storms Dudley, Eunice \& Franklin \\
Mar 2023 & Pluvial & Cork & 1 & Southwest spring-rainfall flooding \\
Sep 2023 & Pluvial & South / east & 1 & Storm Agnes rainfall flooding \\
Oct 2023 & Fluvial & Midleton / East Cork & 2 & Storm Babet: Owenacurra river and surface flooding \\
Nov 2023 & Coastal & South coast & 1 & Storm Ciaran coastal flooding \\
Jan 2024 & Pluvial & North / west & 1 & Storms Isha \& Jocelyn rainfall flooding \\
Apr 2024 & Pluvial & South / west & 1 & Storm Kathleen rainfall flooding \\
Nov 2024 & Pluvial & Cork / Galway & 1 & Storms Bert \& Conall rainfall flooding \\
Dec 2024 & Coastal & West \& south coast & 1 & Storm Darragh coastal flooding \\
Jan 2025 & Pluvial & Galway / Mayo / Donegal & 1 & Storm \'Eowyn rainfall and wind-driven flooding \\
Oct 2025 & Pluvial & South / east & 1 & Storm Amy rainfall flooding \\
Nov 2025 & Fluvial & Wexford / Laois / Portarlington & 1 & Storm Claudia river flooding \\
Dec 2025 & Coastal & Donegal & 1 & Storm Bram marine / coastal warning \\
\bottomrule
\end{longtable}}

\newpage
\section{Annual flood-related wellbeing scores (2012--2025)}\label{app:wellbeing}

\begin{center}

\captionof{table}{Per-indicator annual flood-related wellbeing scores. Each cell reports the mean$\,\pm\,$standard deviation and, in parentheses, the number of substantive discourse records with an indicator score greater than $0$ in that year (a score of $0$ denotes no evidence and is excluded). A dash (--) denotes no non-zero observations.}
\vspace{1em}

\label{tab:appendix_wellbeing}
\resizebox{\textwidth}{!}{%
\begin{tabular}{@{\hspace{2\tabcolsep}}l *{3}{c} *{3}{c} *{3}{c}@{\hspace{2\tabcolsep}}}
\toprule
& \multicolumn{3}{c}{\textbf{BE$-$}} & \multicolumn{3}{c}{\textbf{DO$-$}} & \multicolumn{3}{c}{\textbf{CONNECT$+$}} \\
\cmidrule(lr){2-4}\cmidrule(lr){5-7}\cmidrule(lr){8-10}
\textbf{Year} & I1 & I2 & I3 & I4 & I5 & I6 & I7 & I8 & I9 \\ \midrule
2012 & 1.12$\pm$0.33~(16) & 1.45$\pm$0.50~(31) & -- & 1.00$\pm$0.00~(13) & 1.42$\pm$0.49~(12) & 1.60$\pm$0.49~(5) & 2.00$\pm$0.00~(1) & 1.25$\pm$0.43~(4) & 1.36$\pm$0.48~(33) \\
2013 & 1.50$\pm$0.50~(2) & 1.42$\pm$0.49~(12) & -- & 1.00$\pm$0.00~(2) & 2.11$\pm$0.57~(9) & 1.00$\pm$0.00~(1) & 1.50$\pm$0.50~(2) & 1.00$\pm$0.00~(6) & 1.42$\pm$0.49~(12) \\
2014 & 1.16$\pm$0.37~(44) & 1.09$\pm$0.29~(120) & 1.00$\pm$0.00~(3) & 1.00$\pm$0.00~(21) & 1.10$\pm$0.30~(51) & 1.02$\pm$0.15~(44) & 1.07$\pm$0.26~(28) & 1.07$\pm$0.25~(46) & 1.10$\pm$0.31~(125) \\
2015 & 1.47$\pm$0.50~(30) & 1.32$\pm$0.47~(84) & 2.33$\pm$0.75~(12) & 1.31$\pm$0.46~(16) & 1.58$\pm$0.71~(45) & 1.60$\pm$0.80~(30) & 1.50$\pm$0.50~(32) & 1.39$\pm$0.67~(57) & 1.28$\pm$0.45~(95) \\
2016 & 1.35$\pm$0.48~(26) & 1.16$\pm$0.37~(44) & 1.50$\pm$0.50~(2) & 1.36$\pm$0.48~(11) & 1.48$\pm$0.50~(31) & 1.42$\pm$0.49~(24) & 1.26$\pm$0.44~(27) & 1.27$\pm$0.45~(44) & 1.21$\pm$0.41~(53) \\
2017 & 1.13$\pm$0.34~(23) & 1.25$\pm$0.43~(60) & 1.38$\pm$0.48~(8) & 1.00$\pm$0.00~(10) & 1.12$\pm$0.32~(26) & 1.04$\pm$0.20~(25) & 1.23$\pm$0.42~(13) & 1.29$\pm$0.67~(17) & 1.28$\pm$0.45~(60) \\
2018 & 1.33$\pm$0.47~(9) & 1.09$\pm$0.28~(35) & 1.00$\pm$0.00~(2) & 1.00$\pm$0.00~(2) & 1.29$\pm$0.46~(17) & 1.22$\pm$0.42~(9) & 1.12$\pm$0.33~(8) & 1.00$\pm$0.00~(12) & 1.09$\pm$0.29~(44) \\
2019 & 1.27$\pm$0.45~(11) & 1.29$\pm$0.46~(85) & 1.50$\pm$0.50~(6) & 1.60$\pm$0.49~(5) & 1.22$\pm$0.42~(18) & 1.18$\pm$0.38~(17) & 1.23$\pm$0.42~(13) & 1.14$\pm$0.34~(22) & 1.25$\pm$0.43~(87) \\
2020 & 1.31$\pm$0.46~(70) & 1.35$\pm$0.48~(207) & 1.32$\pm$0.46~(19) & 1.12$\pm$0.33~(58) & 1.39$\pm$0.49~(116) & 1.25$\pm$0.43~(72) & 1.26$\pm$0.44~(72) & 1.18$\pm$0.38~(105) & 1.13$\pm$0.33~(222) \\
2021 & 1.12$\pm$0.32~(26) & 1.20$\pm$0.40~(122) & 1.40$\pm$0.49~(5) & 1.12$\pm$0.33~(16) & 1.17$\pm$0.38~(47) & 1.08$\pm$0.27~(26) & 1.00$\pm$0.00~(33) & 1.02$\pm$0.15~(83) & 1.09$\pm$0.29~(145) \\
2022 & 1.07$\pm$0.25~(15) & 1.04$\pm$0.21~(67) & 1.00$\pm$0.00~(4) & 1.00$\pm$0.00~(10) & 1.03$\pm$0.17~(32) & 1.00$\pm$0.00~(32) & 1.04$\pm$0.20~(25) & 1.02$\pm$0.14~(49) & 1.08$\pm$0.27~(77) \\
2023 & 1.46$\pm$0.56~(119) & 1.41$\pm$0.49~(175) & 1.33$\pm$0.47~(36) & 1.49$\pm$0.50~(45) & 1.50$\pm$0.50~(131) & 1.49$\pm$0.50~(74) & 1.52$\pm$0.50~(60) & 1.38$\pm$0.48~(85) & 1.31$\pm$0.46~(163) \\
2024 & 1.30$\pm$0.46~(99) & 1.35$\pm$0.48~(445) & 1.27$\pm$0.45~(11) & 1.39$\pm$0.49~(70) & 1.43$\pm$0.50~(167) & 1.36$\pm$0.48~(126) & 1.12$\pm$0.32~(77) & 1.13$\pm$0.34~(232) & 1.16$\pm$0.36~(490) \\
2025 & 1.47$\pm$0.56~(116) & 1.20$\pm$0.40~(593) & 1.30$\pm$0.56~(40) & 1.38$\pm$0.48~(72) & 1.35$\pm$0.53~(193) & 1.38$\pm$0.49~(112) & 1.32$\pm$0.47~(97) & 1.11$\pm$0.32~(231) & 1.14$\pm$0.34~(587) \\
\bottomrule
\end{tabular}}
\end{center}

\vspace{1em}

\begin{center}

\captionof{table}{Aggregated annual domain scores. The per-record domain score is the sum of the three indicator scores within the domain; each cell reports the mean$\,\pm\,$standard deviation of this domain score and, in parentheses, the number of records with a non-zero domain sum in that year.}
\vspace{1em}
\label{tab:appendix_wellbeing_domains}
\resizebox{0.45\textwidth}{!}{%
\begin{tabular}{@{\hspace{3\tabcolsep}}l c c c@{\hspace{3\tabcolsep}}}
\toprule
\textbf{Year} & \textbf{BE$-$} & \textbf{DO$-$} & \textbf{CONNECT$+$} \\ \midrule
2012 & 2.03$\pm$0.97~(31) & 1.52$\pm$0.98~(25) & 1.53$\pm$0.92~(34) \\
2013 & 1.67$\pm$0.62~(12) & 2.20$\pm$0.60~(10) & 2.00$\pm$1.04~(13) \\
2014 & 1.45$\pm$0.62~(128) & 1.91$\pm$0.80~(64) & 1.67$\pm$0.92~(130) \\
2015 & 2.13$\pm$1.76~(86) & 3.04$\pm$2.06~(46) & 2.39$\pm$1.78~(104) \\
2016 & 1.78$\pm$1.03~(50) & 2.71$\pm$1.63~(35) & 2.37$\pm$1.47~(65) \\
2017 & 1.70$\pm$0.85~(66) & 2.32$\pm$0.80~(28) & 1.83$\pm$1.28~(63) \\
2018 & 1.41$\pm$0.79~(37) & 1.84$\pm$0.99~(19) & 1.53$\pm$0.83~(45) \\
2019 & 1.49$\pm$0.96~(89) & 2.00$\pm$1.55~(25) & 1.70$\pm$1.05~(88) \\
2020 & 1.84$\pm$1.04~(215) & 2.61$\pm$1.29~(121) & 2.02$\pm$1.27~(230) \\
2021 & 1.44$\pm$0.81~(126) & 1.84$\pm$0.80~(55) & 1.82$\pm$0.76~(152) \\
2022 & 1.27$\pm$0.60~(71) & 1.88$\pm$0.75~(40) & 1.92$\pm$0.92~(83) \\
2023 & 2.30$\pm$1.54~(204) & 2.74$\pm$1.62~(136) & 2.31$\pm$1.49~(182) \\
2024 & 1.65$\pm$0.85~(452) & 2.83$\pm$1.69~(179) & 1.82$\pm$0.96~(503) \\
2025 & 1.50$\pm$0.97~(622) & 2.52$\pm$1.61~(204) & 1.65$\pm$1.04~(637) \\
\bottomrule
\end{tabular}}
\end{center}

\end{document}